\documentclass[]{aa}  

\usepackage{graphicx}
\usepackage{txfonts}

\usepackage{natbib}
\usepackage{hyperref}

\bibpunct{(}{)}{;}{a}{}{,} 

\usepackage{xcolor}
\usepackage{verbatim}

\usepackage{listings}
\usepackage{hyperref}
\usepackage{cleveref}
\usepackage{mathrsfs}
\usepackage{dsfont}
\usepackage{amsmath}
\usepackage{graphicx}
\usepackage{subcaption}

\begin{document}

   \title{Search for quasi-periodic signals in magnetar giant flares}

   \subtitle{Bayesian inspection of SGR 1806-20 and SGR 1900+14}

   \author{Daniel Pumpe
                \and Michael Gabler
                \and Theo Steininger
                \and Torsten A. En{\ss}lin}

   \institute{Max Planck Institute for Astrophysics, 
              Karl-Schwarzschild-Str. 1, 
              D-85741 Garching, Germany
             }

   \date{Received August, 18th 2017 ; accepted November, 17th 2017}

 
  \abstract   {Quasi-periodic oscillations (QPOs) discovered in the decaying 
tails of giant flares of magnetars are believed to be torsional oscillations of 
neutron stars. These QPOs have a high potential to constrain properties of high-density matter. In search for quasi-periodic signals, we study the light curves 
of the 
giant flares of SGR 1806-20 and SGR 1900+14, with a 
non-parametric Bayesian signal inference method called D$^3$PO.
The D$^3$PO algorithm models the raw photon counts as a continuous flux and 
takes the Poissonian shot noise as well as all instrument effects
into account. 
It reconstructs the logarithmic flux and 
its power spectrum from the data. 
Using this fully noise-aware method, we do not confirm 
previously reported frequency lines at $\nu\gtrsim17\,$Hz because they fall into the 
noise-dominated regime. However, we find two new potential candidates for oscillations at $9.2\,$Hz (SGR 
1806-20) and $7.7\,$Hz (SGR 1900+14). If these are real and the fundamental 
magneto-elastic oscillations of the magnetars, current theoretical models would favour relatively weak 
magnetic fields $\bar B\sim 6\times10^{13} - 3\times10^{14}\,$G (SGR 1806-20) 
and a relatively low shear velocity inside the crust compared to previous findings.}


   \maketitle
%

\section{Introduction}
The discovery of quasi-periodic oscillations (QPOs) in the giant flare of the 
magnetar SGR 1806-20 by \cite{Israel2005} may have been the first detection of 
neutron star oscillations and triggered a wealth of theoretical work 
explaining the reported frequencies. The giant flare was likely caused 
by a large-scale reconnection or an interchange instability of the
magnetic field \citep{Thompson1995}. Large amounts of energy are released as an 
expanding $e^\pm$-pair plasma, observable as the initial spike 
of the giant flare. Parts of this plasma are trapped by the ultra-strong 
magnetic field and form a so-called trapped fireball \citep{Thompson1995}, 
which then slowly evaporates on a timescale of up to a few $100$ seconds.
The QPOs were detected in this decaying tail of the giant flare. Other groups have not only 
confirmed this detection, they even found additional oscillation frequencies in different magnetars: $18$, $26$, $29$, $92$, $150$ , $625$, and 
$1840$ Hz in the giant flare of SGR 1806-20, and $28$, $53$, $84$, and $155$\,Hz in the giant 
flare of SGR 1900+14 \citep[e.g.][]{Strohmayer2005, 
Watts2006,Strohmayer2006,Huppenkothen2014c}. With different methods, more 
oscillation frequencies were found in the giant flare of SGR 1806-20 by 
\cite{Hambaryan2011} $17$, $21$, $36$, $59$, and $116\,$Hz. The number of 
giant flares at the time of writing is limited to three events. The more frequent but 
less energetic bursts of several magnetars have therefore  also
been investigated for 
frequencies, and some candidates were found: $57\,$Hz  in SGR 1806-20 
\citep{Huppenkothen2014b}, 
and at $93$, $127$, and $260\,$Hz in SGR 
J1550-5418 \citep{Huppenkothen2014a}. 

It was soon realized that these frequencies are probably related to 
oscillations of the neutron star, and several groups tried to identify them as 
elastic oscillations of the crust \citep{Duncan1998, 
Messios2001, Strohmayer2005, Piro2005, Sotani2007, Samuelsson2007, Steiner2009, 
Samuelsson2009, Sotani2013, Deibel2014, Sotani2016}, Alfv\'en oscillations 
\citep{Cerda2009, Sotani2008, Colaiuda2009}, or coupled magneto-elastic 
oscillations  \citep{Levin2006,Levin2007,Glampedakis2006b, Gabler2011letter, 
Gabler2012, Colaiuda2011, vanHoven2011, vanHoven2012}.
The theoretical models based on the observed frequencies are very elaborate and 
may be able to constrain properties of high-density matter as found in the 
interior of neutron stars. Some of the models, for instance, require a superfluid 
component in the core of the star \citep{vanHoven2011, 
vanHoven2012, Glampedakis2011a, Passamonti2013, Gabler2013b, Passamonti2014, 
Gabler2016, Gabler2017}. Different models depend sensitively on the 
identification of the fundamental oscillation frequency, and may not explain all 
of the observed frequencies. 
Even when the fundamental frequency is identified, the interpretation and 
parameter estimation is not yet straightforward because of degeneracies in the 
parameter space. However, keeping other stellar parameters fixed, some general 
trends of the fundamental oscillation frequency can be summarized as 
follows \citep[see][for a detailed discussion]{Gabler2016,Gabler2017}: i) The frequency
scales linearly with the magnetic field strength. ii) It decreases with 
increasing compactness \citep{Sotani2008}. The compactness is related to the 
hardness of the equation of state (EOS): Material with a stiff equation of 
state is harder to compress, leading to larger radii and hence 
lower compactnesses.
iii) It can only reach the surface for significantly strong magnetic fields 
$\bar B\gtrsim\bar B_\text{outbreak}(\sqrt{c_s})$, whose thresholds depend on the square 
root of the shear velocity \citep{Gabler2017}. 

It is of great importance to understand which of the frequencies are 
in the signal. In previous attempts at identifying possible frequencies in the 
light curve, the statistical noise of the detectors was not modelled consistently.
To improve on this, we employ a Bayesian method that properly takes the 
photon shot noise as the largest contributor into account.

In this work, we re-analyse the data for the two giant flares of SGR1806-20 and SGR1900+14, which were
obtained with the proportional counter array (PCA) of the Rossi X-ray Timing 
Explorer (RXTE). The data are available online at the High Energy 
Astrophysics Science Archive Research Center (HEASARC).
In \Cref{sec:inference} we briefly discuss our numerical approach to estimating 
the influence of the noise and to reconstructing the likely signal with a reduced 
contribution from the noise. The next section is devoted to the reconstruction
and investigation of the light curves of the two giant flares of
SGR 1806-20 and SGR 1900+14, which are then discussed in \Cref{sec:discussion}. 
We conclude our analysis in \Cref{sec:conclusions}.

\section{Inference of photon observations}\label{sec:inference}
Our goal is to reconstruct the light curve and possible frequencies of QPOs in 
giant X-ray flares for the neutron stars SGR 1806-20 and SGR 1900+14 using 
the data taken by RXTE. 
Owing to experimental constraints such as limited data storage, finite 
time resolution of each photon count, finite detector size, and sensitivity, RXTE 
cannot record all physical relevant and available information of a continuous 
photon flux. Most importantly, the recorded photon counts contain significant
photon shot noise. Hence, we have to use probabilisitic data 
analysis methods to obtain an estimate of the time-dependent, continuous photon 
flux $\vec{\phi}\left(t\right)$ , including its uncertainties. 
Since the flux varies on a logarithmic scale, we reconstructed the logarithm of the flux $s\left(t\right) = \log \left(\vec{\phi}\left(t\right) / \phi_0\right)$ 
and its temporal power spectrum directly from the data. 

In this context, it is natural to build the inference upon the
Bayes theorem, 
allowing us to investigate the a posteriori probability distribution 
$P\left(\vec{\phi}\vert \vec{d}\right)$, stating how likely a given photon flux 
field $\vec{\phi}$ is given the data set $\vec{d}$ . 
For the two data sets of SGR 1806-20 and SGR 1900-14, we assumed the data $\vec{d}$ 
to be the result of a Poisson process whose expectation value $\vec{\lambda}$ is 
given by the photon flux $\vec{\phi}\left(t\right)$ of the photon burst 
convolved with the response operator $\vec{R}$,  
\begin{align}
\vec{\lambda} = \vec{R} \vec{\phi}\left(t\right)= \vec{R} \phi_0 e^{s\left(t\right)}\,.
\label{eq:lambda}
\end{align}
The response operator encodes all instrument specifications and states how 
$\vec{\phi}$ imprints itself on $\vec{\lambda}$. Here, we only considered the 
readout deadtime periods of the instrument and neglected all other 
instrument responses.
In \Cref{eq:lambda} the constant $\phi_0$ absorbs
numerical constants and the physical units of the time dependent log- flux signal field $s 
\left(t\right) = \vec{s}$. As the signal describes the logarithmic flux, we 
naturally ensured the positive definiteness of the photon flux. Since we analysed QPOs, it is expected that $\vec{\phi}$ exhibits unknown but spatial
correlations.
Hence, we did not enforce any concrete spatial correlations. We only 
assumed $\vec{\phi}$ to follow a multivariate log-normal statistic, with an 
\textit{\textup{a priori}} unknown covariance. Assuming stationary statistics,
 the underlying covariance is fully determined by a power 
spectrum $P_{s} \left(\nu\right)$, described by $P_{s} \left(\nu\right) = P_0 e^{\vec{\tau}\left(\nu\right)}$, to ensure positivity of the power spectrum. We inferred this power spectrum as well as $\vec{\phi}$ from the data themselves 
by setting up a hierarchical prior model
\citep{Selig:2015ul,2016A&amp;A...586A..76J,2016PhRvE..94a2132P,
2011PhRvD..83j5014E, 2016arXiv161208406E}. 
Hence the posterior of our Bayesian inference is given by
\begin{align}
P\left(\vec{\phi}, \vec{\tau} \vert \vec{d}\right) &\propto \frac{P\left( ,
\vec{d}\vert \vec{\phi},\vec{\tau}\right)P\left(\vec{\tau}\vert 
\sigma_{\mathrm{sm}}\right)}{P\left(\vec{d}\right)}\,,
\end{align}
where $P\left( \vec{d}\vert \vec{\phi},\vec{\tau}\right)$ is the likelihood 
describing how a potential photon flux $\vec{\phi}$ including a certain covariance structure 
described by $\vec{\tau}$ imprints itself in a potential data set. 
In addition to strong spectral peaks in the power 
spectrum induced by the almost discrete frequencies of QPOs, we enforce some 
kind of spectral smoothness on the logarithmic scale, favouring power-law spectra. This behaviour is enforced 
and tuned by the prior term $P\left(\vec{\tau}\vert \sigma_{\mathrm{sm}}\right)$ and its parameter $\sigma_{\mathrm{sm}}$ 
(\cite{2013PhRvE..87c2136O}). If $\sigma_{\mathrm{sm}} \rightarrow 0,$ infinite spectral 
smoothness is enforced, while $\sigma_{\mathrm{sm}} \rightarrow \infty$ does not enforce 
any spectral smoothness. As we wish to be sensitive to spectral lines in the 
power spectrum, the influence of $\sigma_{\mathrm{sm}}$ on the inference is discussed
in greater detail in \Cref{sec:performance}. 

For the detailed mathematical rigorous derivation and discussion of the used D$^3$PO-
algorithm, we refer to \cite{Selig:2015rt} and 
\cite{2016arXiv161208406E}. The D$^3$PO-
algorithm was successfully applied on the Fermi LAT data \citep{Selig:2015ul}. In order to handle high 
power spectrum resolutions as is needed to infer spectral lines in the power spectrum, we reimplemented the 
algorithm in \textsc{NIFTy\,3} \citep{2017arXiv170801073S}. 
The capabilities of the algorithm
to analyse QPOs are demonstrated in the \Cref{sec:performance,sec:injection,sec:QPO_injection}.

\section{Results}
\subsection{SGR 1806-20}
We re-analysed the archival RXTE data of the giant flare of SGR 1806-20, which 
occurred on 2004 December 27. Owing to the high photon flux, the instrument 
telemetry was saturated, causing several data gaps in the first seconds of the 
flare. We therefore neglected the very beginning of the flare in the current analysis 
and started our investigation roughly $4.5\,$s after the initial rise, immediately 
after the third deadtime interval. However, we accounted for the only remaining 
operational down time of the instrument in our data between $ 7.3865 \, 
\text{s} \leq \text{t} \leq 7.9975\,  \text{s}$. For the analysis, we binned 
the data into pixels with a volume of $1/800\,$s. To overcome the 
periodic boundary conditions introduced by the fast-Fourier transformation, which we used to switch between 
signal space and its harmonic space, we performed the signal inference on a 
regular grid with $2^{19}$ pixel, each also with a volume of $1/800\,$s, by adding sufficient buffer time. 

\begin{figure*}[]
\centering      
        \begin{subfigure}{17cm}
        \centering
                        \includegraphics[width=.875\textwidth]{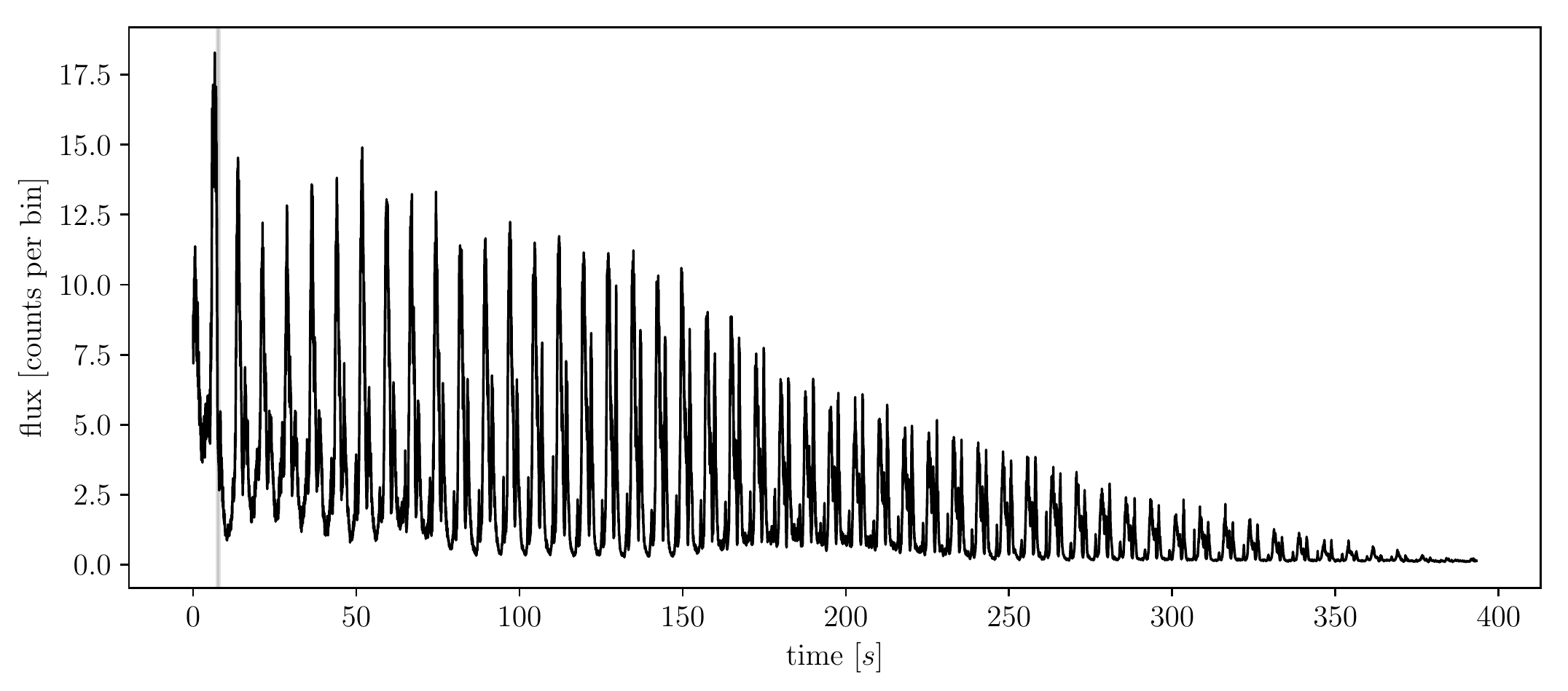} 
                        \caption{}
                        \label{sub_fig:RXTE_whole}
        \end{subfigure}
        \begin{subfigure}{17cm}
        \centering
                        \includegraphics[width=.875\textwidth]{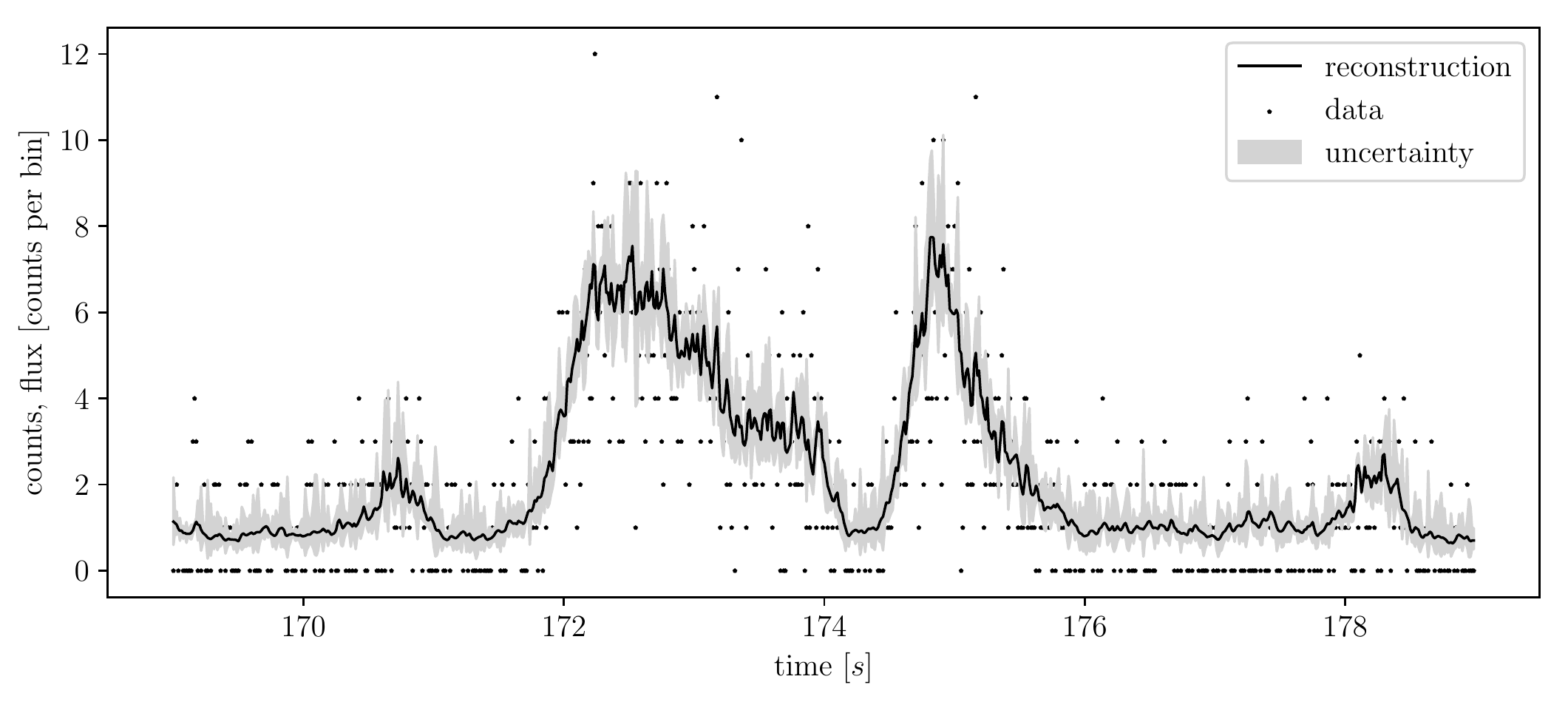} 
                        \caption{}
                        \label{sub_fig:RXTE_snapshot}
        \end{subfigure}
        \caption{Reconstructed light curve of the giant flare of SGR 
1806-20 using a smoothness-enforcing prior, with $\sigma_{\mathrm{sm}} = 5 
\times 10^5$ is shown in \Cref{sub_fig:RXTE_whole}.
The grey narrow rectangle indicates the operational down time of 
the instrument. For better visibility, we plot the light curve between $\approx 
164\,$s and $\approx 174\,$s in \Cref{sub_fig:RXTE_snapshot}. In 
addition to the raw photon counts (black dots), the black line indicates the 
reconstruction of the expected photon counts, i.e. $\lambda$ as well as its one-$\sigma$ confidence interval. Owing to the high resolution of the photon flux, only 
every fourth point of the regular grid is plotted. In both plots each pixel 
has a duration of $1/800\,$s .}
\label{fig:SGR1806_reconstruction}
\end{figure*}

\begin{figure*}
\centering
\includegraphics[width=.875\textwidth]{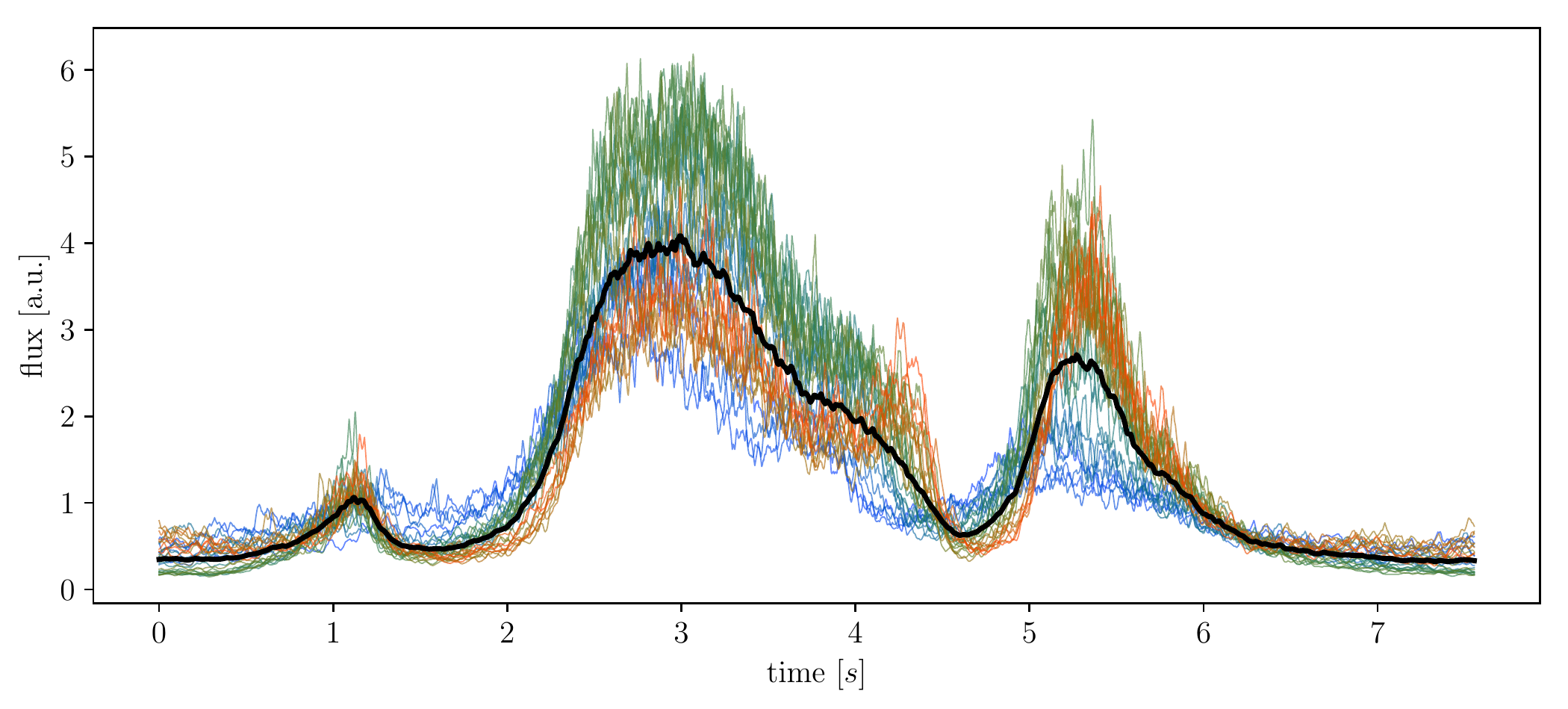}
\caption{Pulse profiles of different rotation periods 
overplotted as given in \Cref{sub_fig:RXTE_whole}. The temporal evolution of the 
pulse profile is visible from blue to green to red. 
The mean pulse is shown as a thick black line. All pulses are shifted such that 
their log-means are zero.}
\label{fig:mean_period_RXTE}
\end{figure*}

\begin{figure*}[]
\begin{subfigure}{17cm}
        \centering
                        
\includegraphics[width=.875\textwidth]{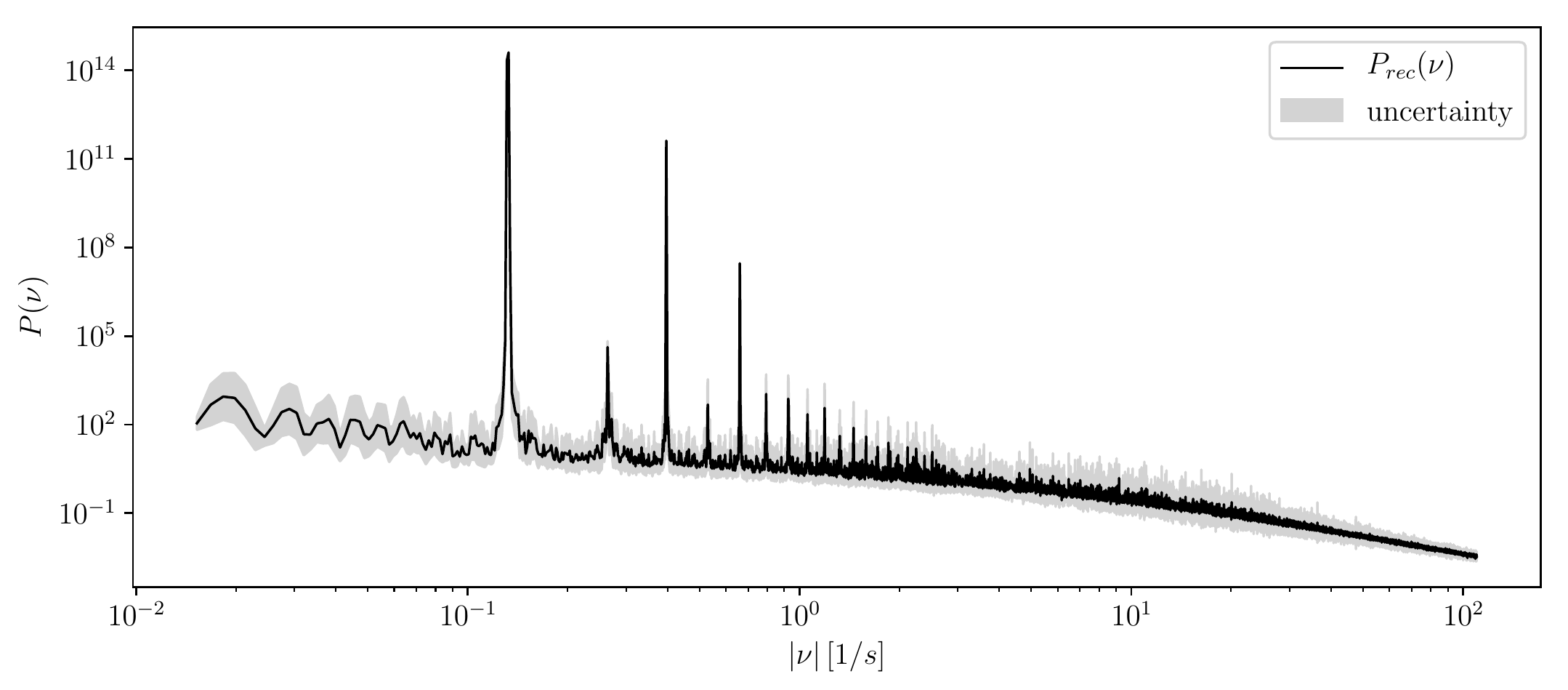} 
                        \caption{}
                        \label{sub_fig:RXTE_spec_5e5}
        \end{subfigure}
\begin{subfigure}{17cm}
        \centering
                        
\includegraphics[width=.875\textwidth]{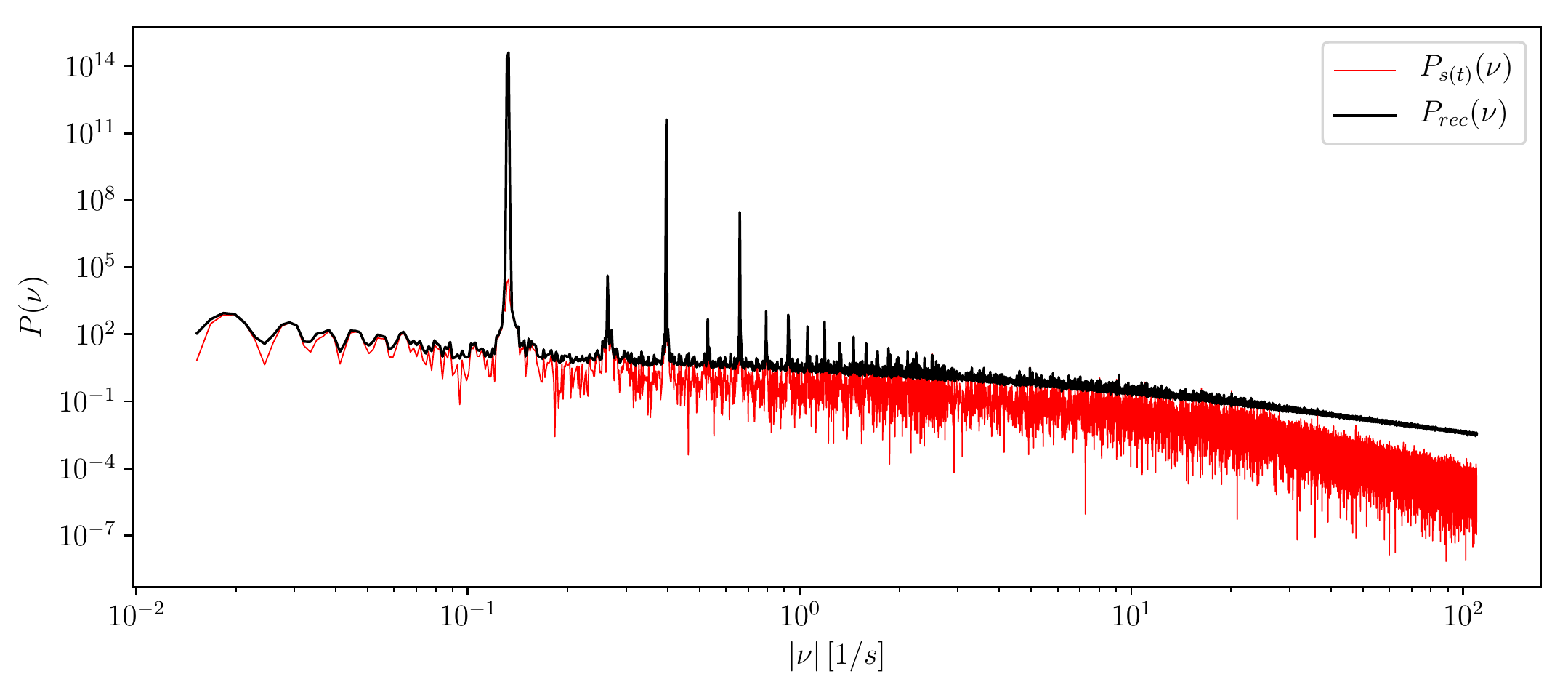} 
                        \caption{}
                        \label{sub_fig:RXTE_combi_spec_5e5}
        \end{subfigure}
        \begin{subfigure}{17cm}
        \centering
\includegraphics[width=.875\textwidth]{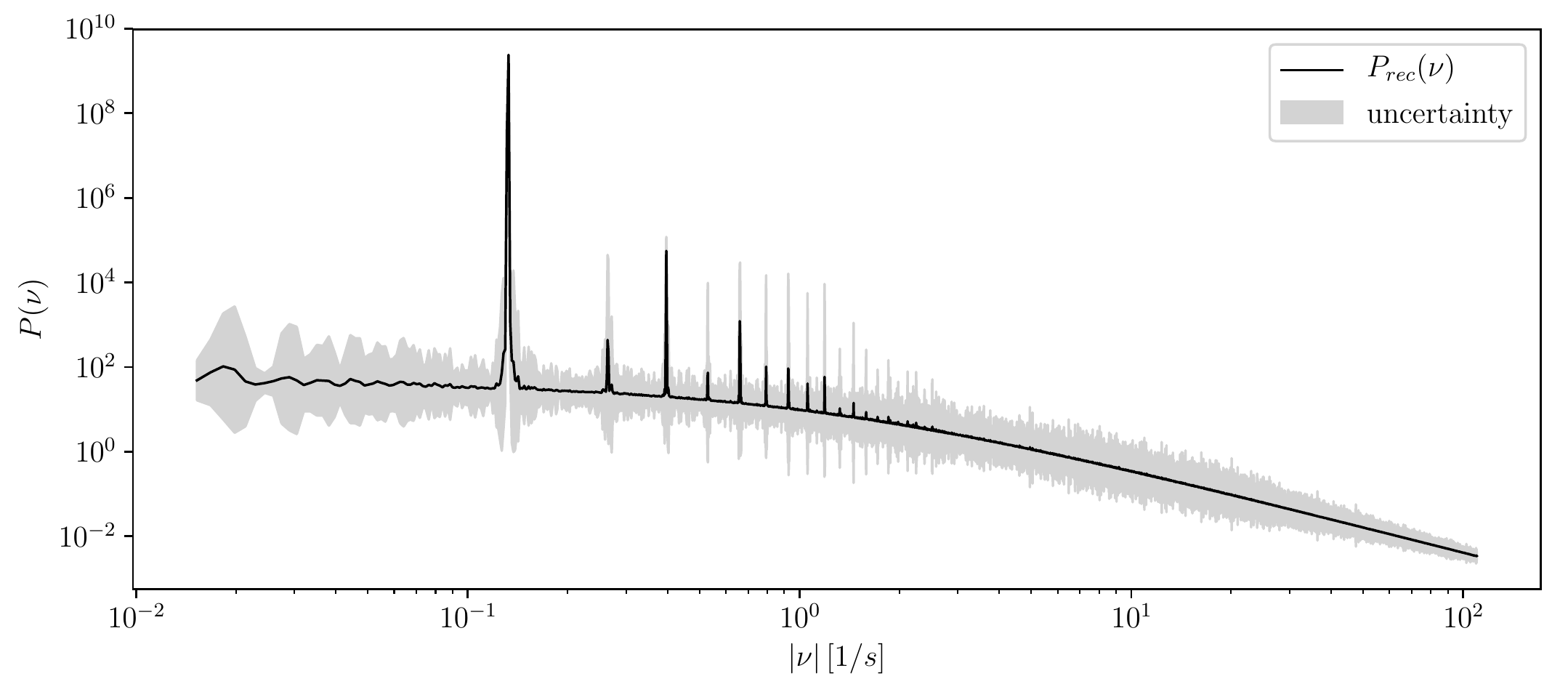} 
                        \caption{}
                        \label{sub_fig:RXTE_spec_1e5}
        \end{subfigure}

    \caption{Reconstructed power spectra of the giant 
flare of SGR 1806-20: 
For \Cref{sub_fig:RXTE_spec_5e5} and \Cref{sub_fig:RXTE_spec_1e5} we used 
smoothness-enforcing priors with $\sigma_{\mathrm{sm}} 
=  5 \times 10^{5}$ and $\sigma_{\mathrm{sm}} =  10^{5}$, respectively.
The uncertainty intervals are given as grey shaded areas. In \Cref{sub_fig:RXTE_combi_spec_5e5}
we show the reconstructed power spectrum again from the data themselves as in \Cref{sub_fig:RXTE_spec_5e5}, along with
the power spectrum of the logarithmic reconstructed light curve of 
\Cref{sub_fig:RXTE_whole}.
Because of the high resolution of the reconstructed power spectra, only every fourth point of 
the regular grid is plotted. Each pixel has a volume of $1/655\,$Hz. }
\label{fig:SGR1806_power}
\end{figure*} 
In \Cref{fig:SGR1806_reconstruction} we plot the inferred $\vec{\phi}$ for the 
entire duration of the giant flare and for a selected period of time 
for a smoothness parameter $\sigma_{\mathrm{sm}} = 5 \times 10^5$. Our algorithm is able 
to significantly reduce the scatter of the light curve that is
caused by the photon shot 
noise. The thus reconstructed light curves can now be further analysed for 
potential periodic signals.

\Cref{fig:mean_period_RXTE} shows the reconstructed profiles of one pulse rotation period.
 There, the mean pulse profile is given as a thick 
black line, and all individual pulses are plotted in different colours. Dark blue are the 
first pulses that have a significantly lower second maxima around $5.5\,$s than 
the mean pulse. The major maximum ($\mathrm{time}\sim3\,$s) is very close to the mean 
maximum. At intermediate times (green lines), both maxima of the pulse 
take their maximum values, roughly $40\%$ and $130\%$ more than at the 
beginning for the first and second maximum, respectively. At late times 
(red lines) the main peak declines by $40\%,$ while the second maximum 
almost stays constant and has an amplitude similar to the main peak. This 
behaviour indicates a complex evolution of the fireball, or of the 
fireballs, if there are more than one. 

The power spectrum of the entire flare is plotted in 
\Cref{sub_fig:RXTE_spec_5e5}.
The rotation period of the magnetar is recovered with the 
frequency of the first main peak
$\nu_0=0.1323\,$Hz, which is very close to $\nu_0=0.13249\,$Hz, as 
given in \cite{2014ApJS..212....6O}.
We are able to find up to the $31$st overtone of this frequency 
at $\nu=4.245\,$Hz. 
In \Cref{sub_fig:RXTE_combi_spec_5e5} we show the reconstructed power spectrum 
$P_\text{rec}$ from \Cref{sub_fig:RXTE_spec_5e5} in black together with the 
power spectrum obtained from the reconstructed light curve 
(\Cref{sub_fig:RXTE_whole}) $P_{s\left(t\right)}\left(\nu\right)$ in red. At 
low frequencies, that is,
$ \nu \lesssim 3\, $[Hz], the two spectra are in good agreement, as the reconstruction is
well constrained by the data on large scales. Between $3\, \text{[Hz]} \lesssim \nu \lesssim 20\, \text{[Hz]}$, 
the inference algorithm enters the regime of a lower signal-to-noise ratio (S/N), which in principle leads to noisier $P_{\mathrm{rec}}$. However, this natural behaviour is counteracted by the smoothness-enforcing prior. 
At higher $\nu \gtrsim 20$ Hz, the shapes of both spectra start to deviate significantly. The reason for this is that in the noise-dominated frequency regime, 
D$^3$PO filters out the photon shot noise. From a naive perspective, small-scale features in the signal therefore need to be significantly strong in order to be detectable after a pure photon shot noise filtering operation on the data set. 
However, D$^3$PO accounts for the power loss of this filtering when it reconstructs the power spectrum from the data themselves.
Thus, \Cref{sub_fig:RXTE_combi_spec_5e5} indicates that above $20$ Hz 
the data are noise dominated, and spectral features there have to be very strong to 
be recognisable. 
To test the dependence of our method on the chosen smoothness prior 
$\sigma_{\mathrm{sm}}$ as discussed also in \Cref{sec:performance}, we additionally 
calculated the reconstructed light curve and its corresponding power spectrum 
for $\sigma_{\mathrm{sm}} =  10^5$. The latter is given in 
\Cref{sub_fig:RXTE_spec_1e5}. Obviously, a smaller $\sigma_{\mathrm{sm}}$ leads to a 
smoothing of the spectrum and the algorithm suppresses the detection of 
periodic signals at higher frequencies. We found $\sigma_{\mathrm{sm}}=5\times10^5$ to 
be the optimal value to still observe power in the Fourier transform at higher 
frequencies. For higher values of $\sigma_{\mathrm{sm}}$ , we qualitatively 
obtain similar but more noisy results for the reconstructed light curve and the 
corresponding power spectra.

\begin{figure}
\centering
                        \includegraphics[width=.5\textwidth]{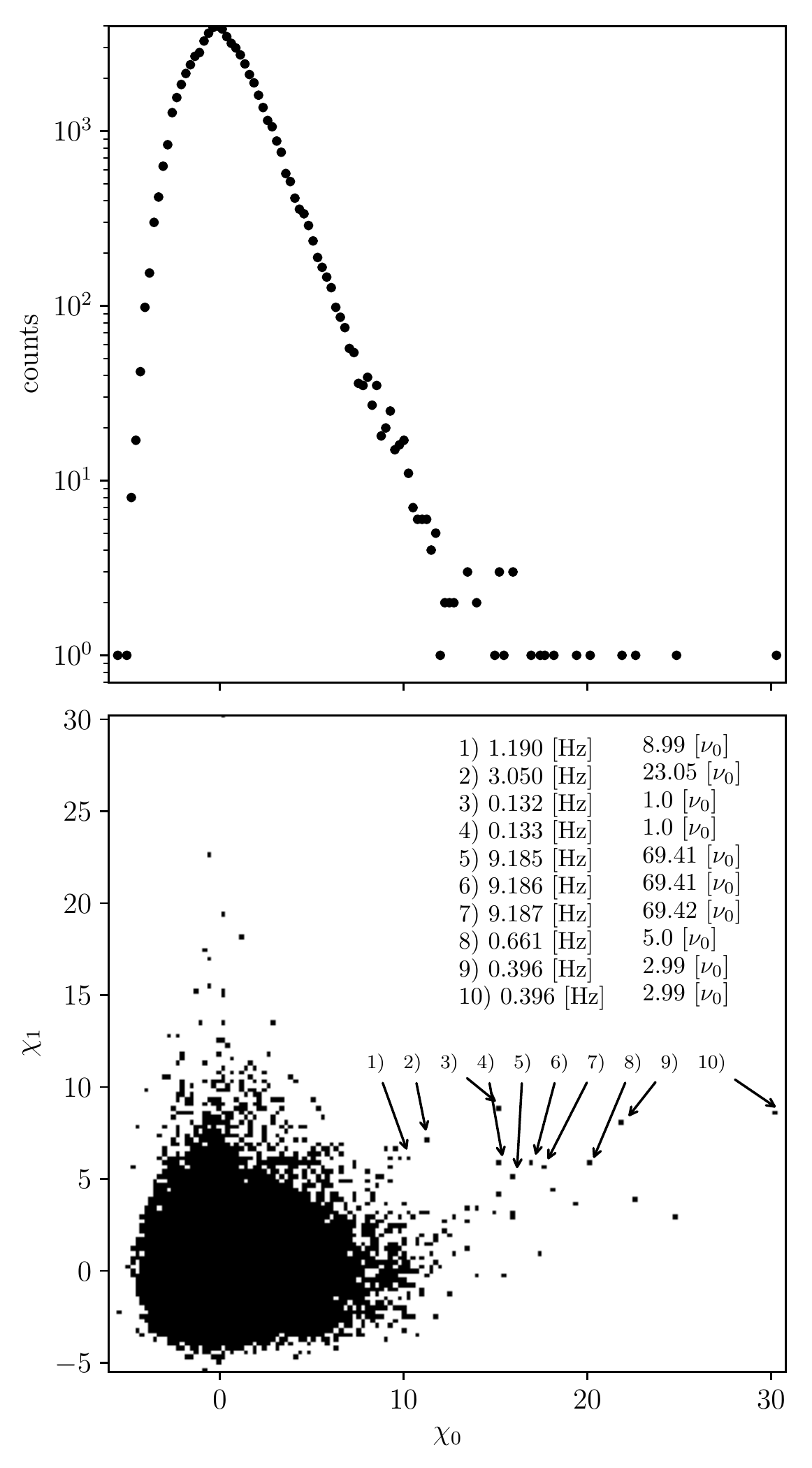}
\caption {Top panel: Histogram of the 
residuals between the logarithmic power spectrum of \Cref{sub_fig:RXTE_spec_5e5} 
and its local median, weighted with its local variance $\sigma$. Bottom 
panel: Two-dimensional histogram of the same quantity 
at some frequency at the x-axis and its first harmonic at the y-axis. In 
addition, we indicate for all counts with $\chi_1\geq 5$ and $\chi_0 \geq 10$ 
the corresponding frequency in Hertz and its multiple of the neutron star 
frequency, $\nu_0 = 0.1323\,$Hz. To generate the histograms, we used the 
power spectrum shown in \Cref{sub_fig:RXTE_spec_5e5}, i.e. $\sigma = 5 \times 10^5$.} 
\label{fig:hists_RXTE}
\end{figure}

In addition to the obvious peaks that are related to the rotation period and the 
corresponding overtones, there are still other features in the reconstructed 
power spectrum of \Cref{sub_fig:RXTE_spec_5e5} that seem to have higher powers than the 
noise. To estimate the significance of these spectral peaks, we calculated a 
residual $\chi$ between the inferred log-spectrum $\tau$ and its local median 
$\bar{\tau}$ weighted with the local variance $\sigma$, 
\begin{align}
\chi = \frac{\tau-\bar{\tau}}{\sigma}\,.
\end{align}
The local median and local variance were calculated over a window of $401$ 
pixels, corresponding to a frequency window of approximately $1\,$Hz. In the top panel of \Cref{fig:hists_RXTE} we plot the 
histogram of 
$\chi_0$, where the index $0$ refers to the fundamental frequency, that is,  $\chi$ at the respective 
frequency. The resulting distribution 
deviates significantly from a Gaussian, 
as there is a significant excess for large $\chi_0$. These counts 
can easily be identified with the highest spectral peaks in 
\Cref{sub_fig:RXTE_spec_5e5} as integer multiples of the neutron star rotation
frequency of $\nu_0=0.1323\,$Hz. 
The fat tails of the distribution make it hard to 
identify whether a peak sticks out of the tail, in particular for 
$\chi_0\gtrsim10$. For $\chi_0\gtrsim15$ all peaks can be identified and are 
related to the neutron star rotation period, except for one at 
$\nu\sim9.187\,$Hz. In \Cref{table:chi_RXTE_10} we show all frequencies that 
have $\chi_0>11$ to have a selection of possible oscillation candidates. If 
there are two or more neighbouring frequencies with $\chi_0>11$, we list 
the highest value.
\begin{table}[]
\caption{All frequencies above $3.5\,$Hz with $\chi_0 > 11$ and 
their multiplicity $n$ of the rotation period $\nu_0 = 0.13249$ [Hz] for SGR 
1806-20.}
\label{table:chi_RXTE_10}
\centering
\begin{tabular}{ccc || ccc}
$\nu$ [Hz] & $\chi_0$ & $n$ [$\nu_0$]& $\nu$ [Hz] & $\chi_0$ & $n$ [$\nu_0$]\\
\hline
3.86&11.071&29.134&11.171&11.762&84.316\\
4.602&11.141&34.735&15.768&11.999&119.013\\
4.95&13.592&37.361&16.272&12.772&122.817\\
6.81&11.212&51.4&19.034&12.632&143.664\\
9.187&18.293&69.341\\

\end{tabular}
\end{table}
All other candidates in \Cref{table:chi_RXTE_10} have significantly lower 
$\chi$ than the oscillation at $9.187\,$Hz and are consistent with being in the 
tail of the distribution that is shown in the top panel of \Cref{fig:hists_RXTE}. This 
indicates that these are artefacts of the Poisson noise.

\begin{figure}
\centering
\includegraphics[width=.5\textwidth]{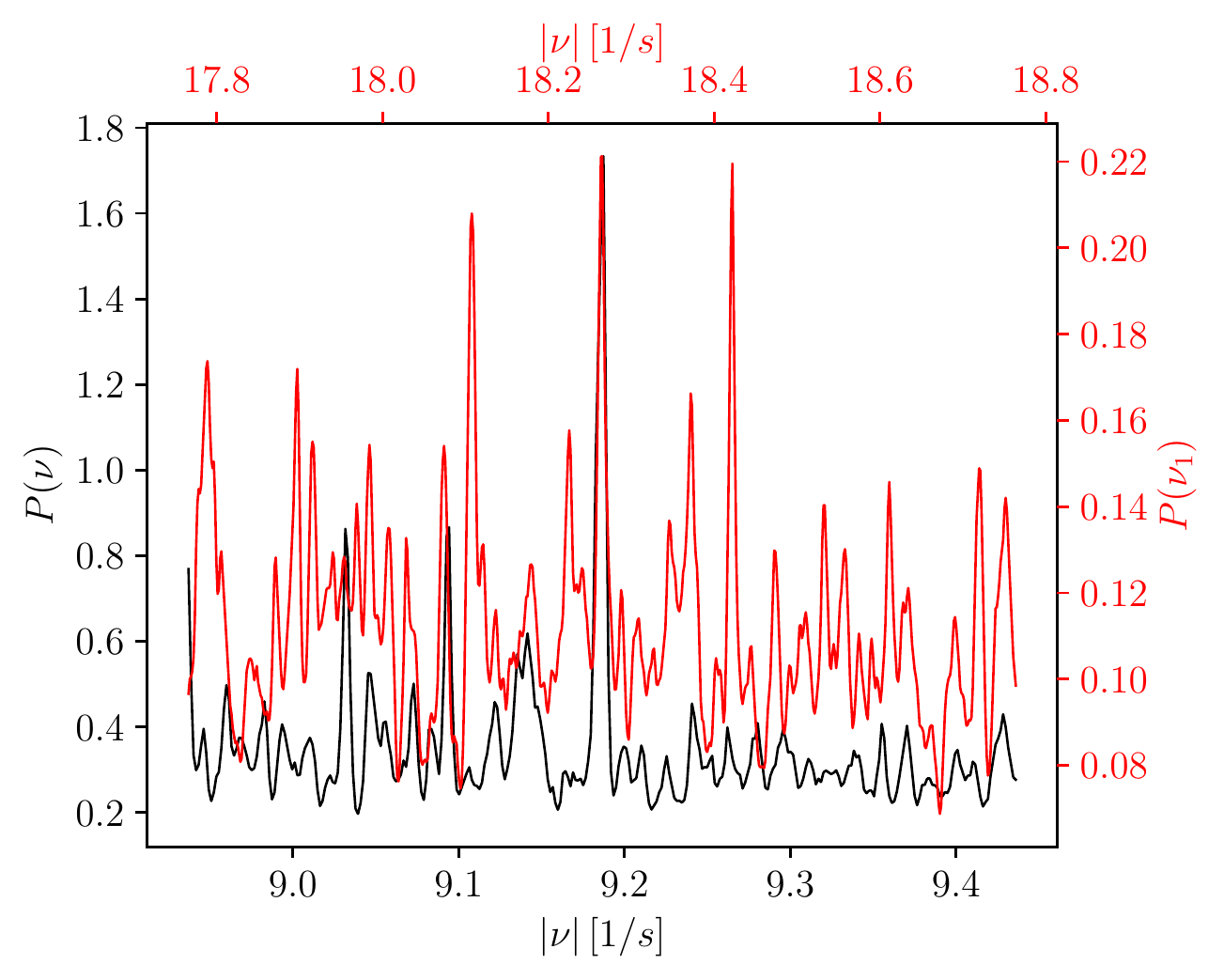}
\caption{Zoomed-in view of the reconstructed power spectra of giant flare SGR 1806-20 around $\nu\sim9.186\,$Hz in black and its first overtone around $\nu\sim18.265\,$Hz in red. The spectra correspond to $\sigma_{\mathrm{sm}} = 5 \times 10^5$ as in \cref{sub_fig:RXTE_spec_5e5}}
\label{fig:zoom_1806_power}
\end{figure}

\begin{table}[]
\caption{Maximum $\chi_0$ at $\nu_\text{max}$ in a $5\%$ interval around 
previously observed oscillation frequencies $\nu$ for SGR 1806-20.
We also show the local variance $\sigma_\chi$ of $\chi_0$ at this interval.}
\label{table:chi_RXTE}
\centering

\begin{tabular}{c c  c c}
$\nu$ [Hz]  & $\nu_{\text{max}}\,$[Hz] in $\pm 5\%$ interval & 
$\chi_{0,\text{max}}$ & 
$\sigma_\chi$\\
\hline 
16.9 & 16.27 & 12.772 & 6.623 \\
18.0 & 18.265 & 10.789 & 5.229\\
21.4 & 20.609 & 9.191& 4.194\\
26.0 & 26.686 & 8.709 & 3.917\\
29.0 & 28.87 & 7.121 & 3.234\\
36.8 & 36.412 & 8.565 &2.82\\
59.0 & 56.592 & 5.146 & 2.258\\
61.3 & 63.162 & 5.027 & 2.114\\
92.5 & 90.364 & 4.699 & 1.859\\
116.3& 118.346 & 4.363 &1.817\\
150  & 152.252 & 4.316 &1.675 
\end{tabular}
\end{table}
We also checked the $\chi_0$ values of previously reported frequencies. None of 
them reach more than $\chi_0\gtrsim5$. We therefor extended our search in a 
$\pm5\%$ interval around the frequencies in \Cref{table:chi_RXTE}. The only frequency higher than  $\chi_0=11$ is at $\nu=16.27\,$Hz. Thus all reported lines are 
consistent with noise.
However, we already see an interesting pattern emerging: In \Cref{table:chi_RXTE} we
locally (within the $\pm5\%$ interval) find the highest powers at $18.265$ and 
$36.412\,$Hz. These are almost twice and four times the only 
significant frequency at $\nu=9.187\,$Hz that our method detects beyond the rotational frequency 
and its first 31 harmonics. In \Cref{fig:zoom_1806_power} we plot 
the reconstructed power spectrum around the $\nu =9.187$ Hz candidate 
oscillation (black line) and its first overtone (red line). The amplitude at 
$\nu =9.187$ Hz is significantly larger (factor 2) than other amplitudes in the 
shown frequency range, while the amplitude at $\nu = 18.265\,$Hz is comparable 
with other spectral peaks.
As discrete frequencies in a power 
spectrum are likely to show spectral peaks at integer multiplies of a ground 
frequency, we also display the two-dimensional histogram of the calculated weighted 
residual $\chi_0$ at some ground frequency on the x-axis and its first 
harmonics $\chi_1$ on the 
y-axis in the bottom panel of \Cref{fig:hists_RXTE}. We marked all 
counts with $\chi_1 \geq 5$ and $\chi_0 \geq 10$ with their corresponding 
frequency in Hz. 
We find about $\text{ten}$ frequencies that satisfy this criterium. Obviously, all but 
the frequencies around $\nu\sim9.186\,$Hz are integer multiples of the rotation 
frequency $\nu_0$. 

This further increases our  confidence that $\nu\sim9.186\,$Hz is a candidate for an 
additional periodic signal in the data. We do not find any significant features 
for frequencies higher than the corresponding overtones that the algorithm 
recovered at $\nu\sim18.265\,$Hz and $\nu\sim36.412\,$Hz.

\subsection{SGR 1900+14}

Analogously to SGR 1806-20, we also re-analysed the archival RXTE data of the 
giant flare of SGR 1900+14 that occurred on 1998 August 27. The resolution of the 
signal field is again $1/800\,$s, but here we only used  $2^{18}$ pixels, as 
the giant flare did not last as long as that of SGR 1806-20. The operational 
down times\footnote{At the following intervals in seconds, the instrument did not 
record any data: $1.89 \leq \text{t} \leq 15.12; 20.84625 \leq \text{t} \leq 
31.1175; 37.54125 \leq \text{t} \leq 47.11875; 55.405 \leq \text{t} \leq 63.12; 
71.87875 \leq \text{t} \leq 79.12; 88.245 \leq \text{t} \leq 95.1375; 106.86 
\leq \text{t} \leq 111.12; 124.285 \leq \text{t} \leq 127.1225; 141.935 \leq 
\text{t} \leq 143.1225$} of RXTE during the observation were taken into 
account for the inference. 

\begin{figure*}[]
\centering
        \begin{subfigure}{17cm}
        \centering
                        \includegraphics[width=.875\textwidth]{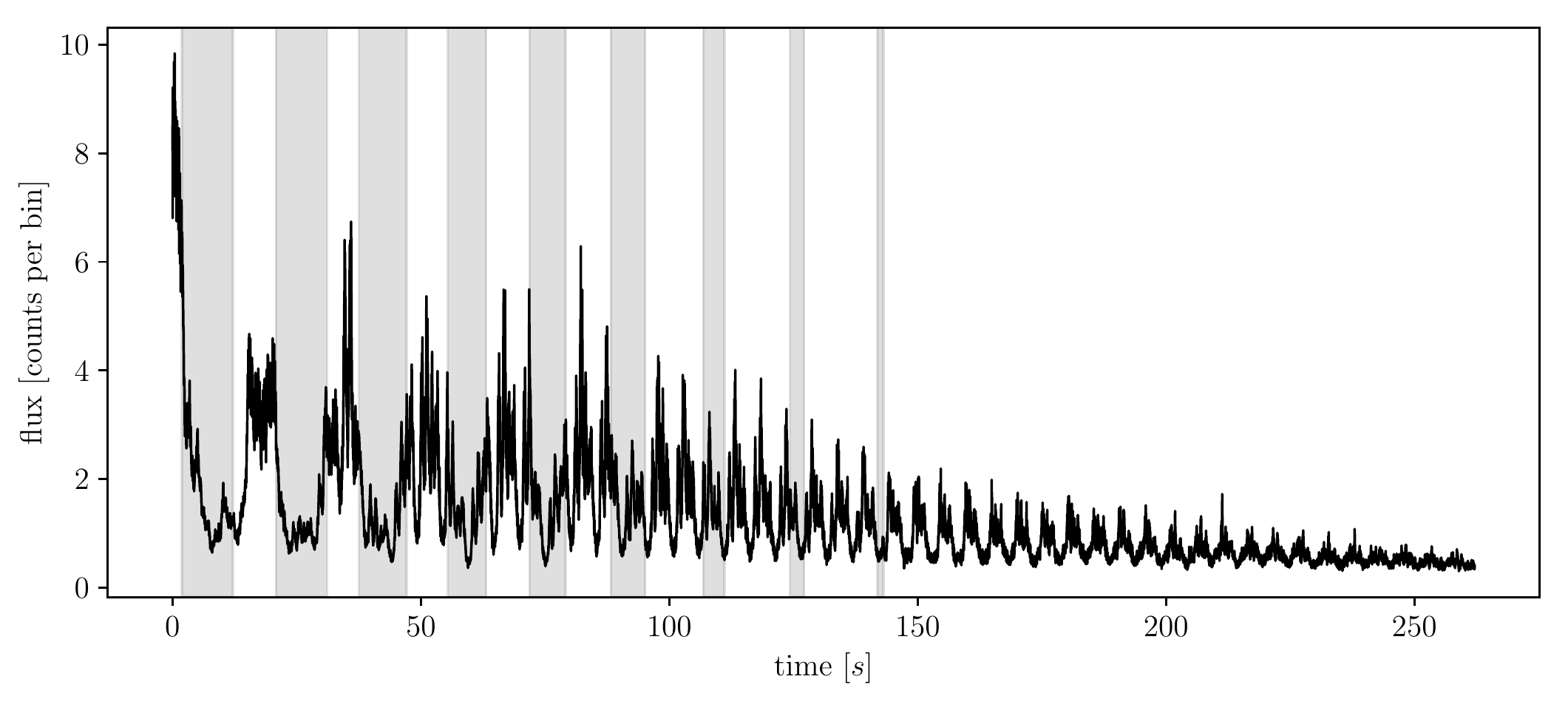}
                        \caption{}
                        \label{sub_fig:SGR_whole}
        \end{subfigure}
        \begin{subfigure}{17cm}
        \centering
                        \includegraphics[width=.875\textwidth]{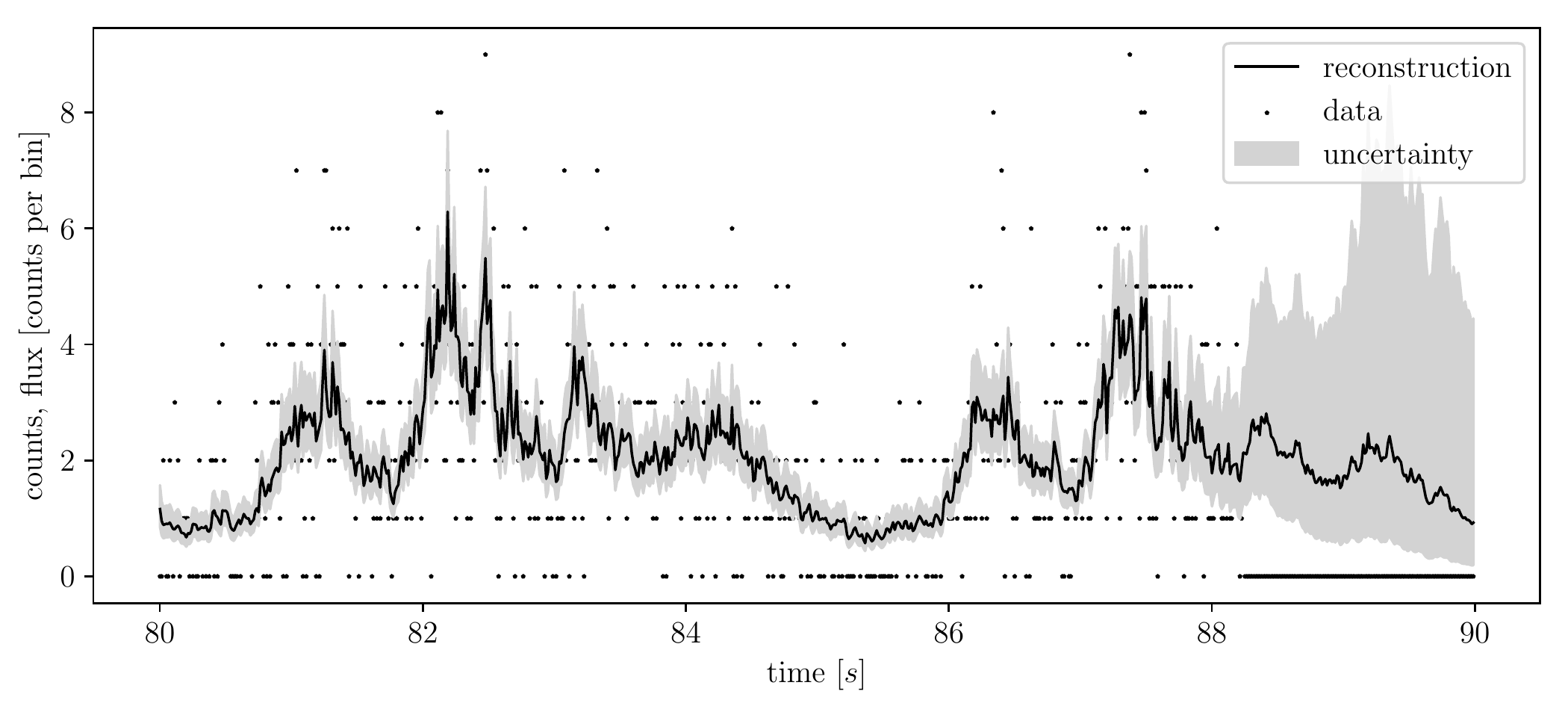}
                        \caption{}
                        \label{sub_fig:SGR_snapshot}
        \end{subfigure}
        \begin{subfigure}{17cm}
        \centering
                        \includegraphics[width=.875\textwidth]{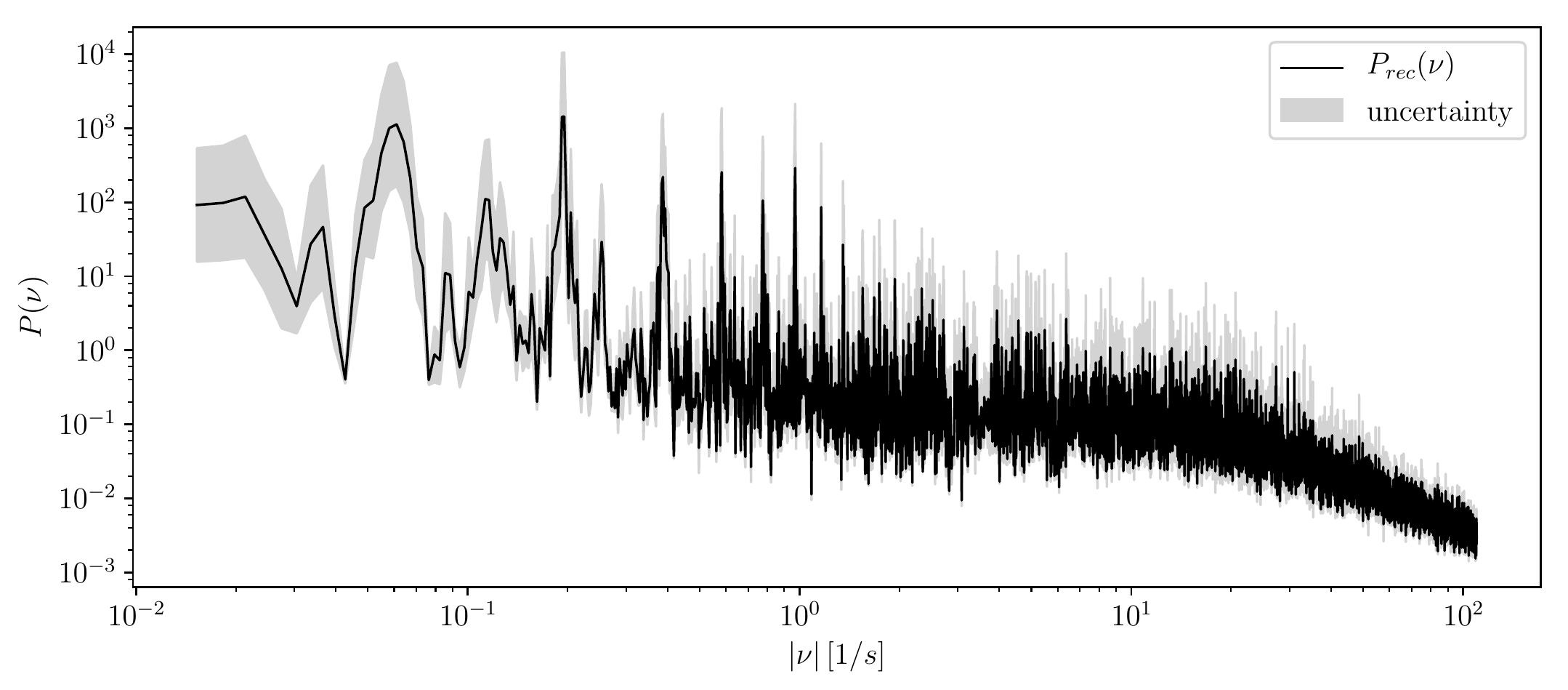}
                        \caption{}
                        \label{sub_fig:SGR_spec_5e5}
        \end{subfigure}
 \end{figure*}
 \begin{figure*}[]
 \ContinuedFloat
        \begin{subfigure}{17cm}
        \centering
                        \includegraphics[width=.875\textwidth]{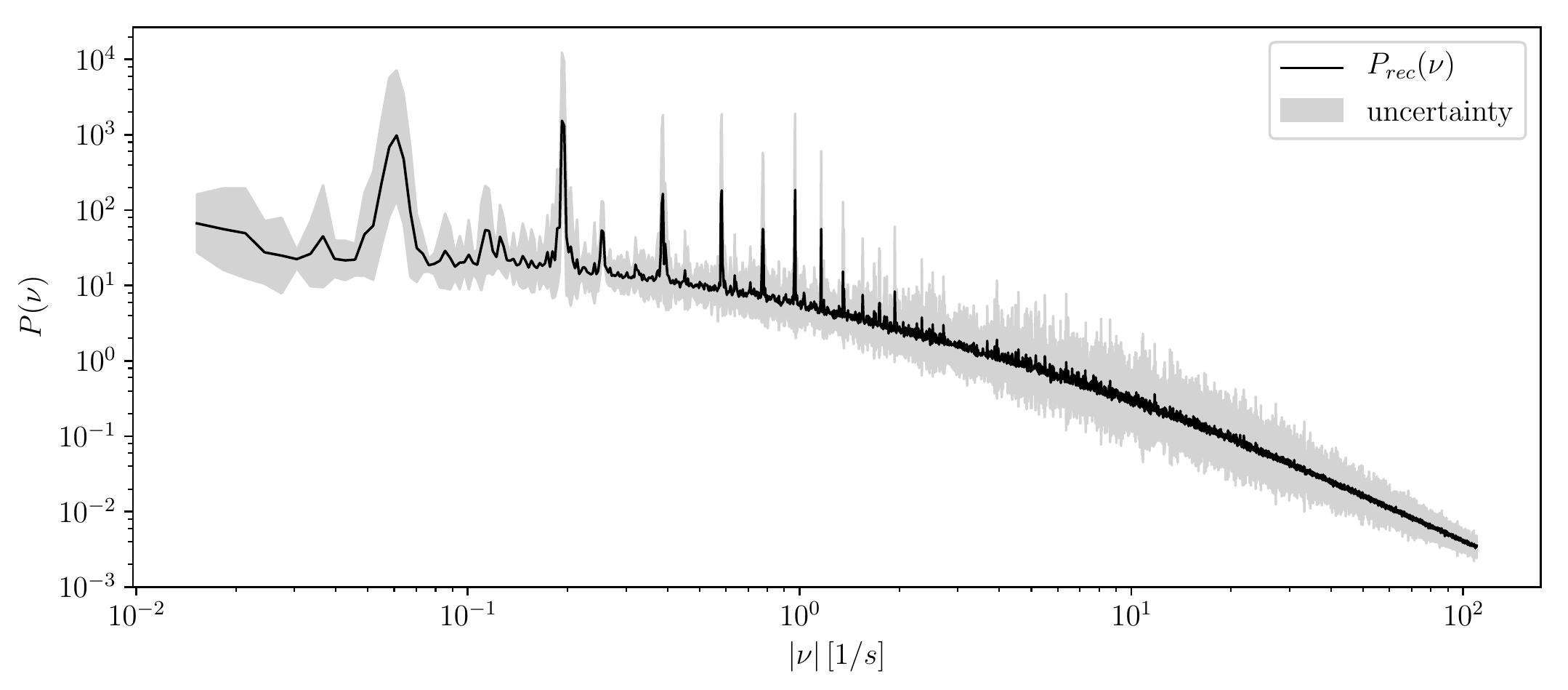}
                        \caption{}
                        \label{sub_fig:SGR_spec_1e5}
        \end{subfigure}
    \caption{\Cref{sub_fig:SGR_whole} shows the reconstructed light curve $\vec{\phi}$ for 
SGR 1900+14. The grey rectangles indicate the operational down times of the 
instrument. 
 \Cref{sub_fig:SGR_snapshot} shows a snapshot of the entire 
 reconstruction between $\approx 80$ sec and $\approx 91$ sec. For $t \gtrsim 88 
 $ the instrument had an operational down time, leading to zero counts in this 
 time interval. 
\Cref{sub_fig:SGR_spec_5e5} 
and \Cref{sub_fig:SGR_spec_1e5}
shows the reconstructed power spectra as well as their uncertainties, using a 
smoothness-enforcing prior with 
$\sigma_{\mathrm{sm}} = 5 \times 10^{5}$
and  $\sigma_{\mathrm{sm}}= 10^{5}$, respectively. 
The plots have the same volumes and sampling rate as in 
\Cref{fig:SGR1806_reconstruction}.}
\label{fig:SGR_reconstruction}
\end{figure*} 

As for SGR1806-20, we were able to recover the frequency corresponding 
to the rotation as $\nu_0=0.1938\,$Hz, which is consistent with the reported 
$\nu_0=0.1923\,$Hz \citep{2014ApJS..212....6O}. For SGR 1900+14, the 
corresponding spectrum with the same smoothness prior $\sigma_{\mathrm{sm}}=5\times10^5$ 
is more noisy because of the various operational down times of the instrument, which are marked grey in 
\Cref{sub_fig:SGR_whole}. In \Cref{sub_fig:SGR_snapshot}, we show a snapshot of 
the entire reconstructed 
light curve for a more detailed view. Even though no data were recorded 
between $88.245\, \text{s}\leq \text{t} \leq 95.1375\, \text{s,}$ the algorithm was able to infer
a light curve by extrapolating from the data-constrained regions. This extrapolation 
 is mainly driven by periodic features appearing in the data. As a further natural consequence, the one-$\sigma$ confidence 
 interval increases drastically during down times.
In total, these multiple operational down times of the instrument lead to fewer observed photons 
and in consequence to a weaker constrained reconstruction of 
the light curve as well as its power spectrum.  Since a smoothness-enforcing prior
$\sigma_{\mathrm{sm}}= 5\times 10^5$ does not sufficiently denoise the light curve, 
we used a slightly stronger prior, $\sigma_{\mathrm{sm}}=  10^5$ 
(\Cref{sub_fig:SGR_spec_1e5}), for our further analysis.

\begin{figure}
\centering
\includegraphics[width=.5\textwidth]{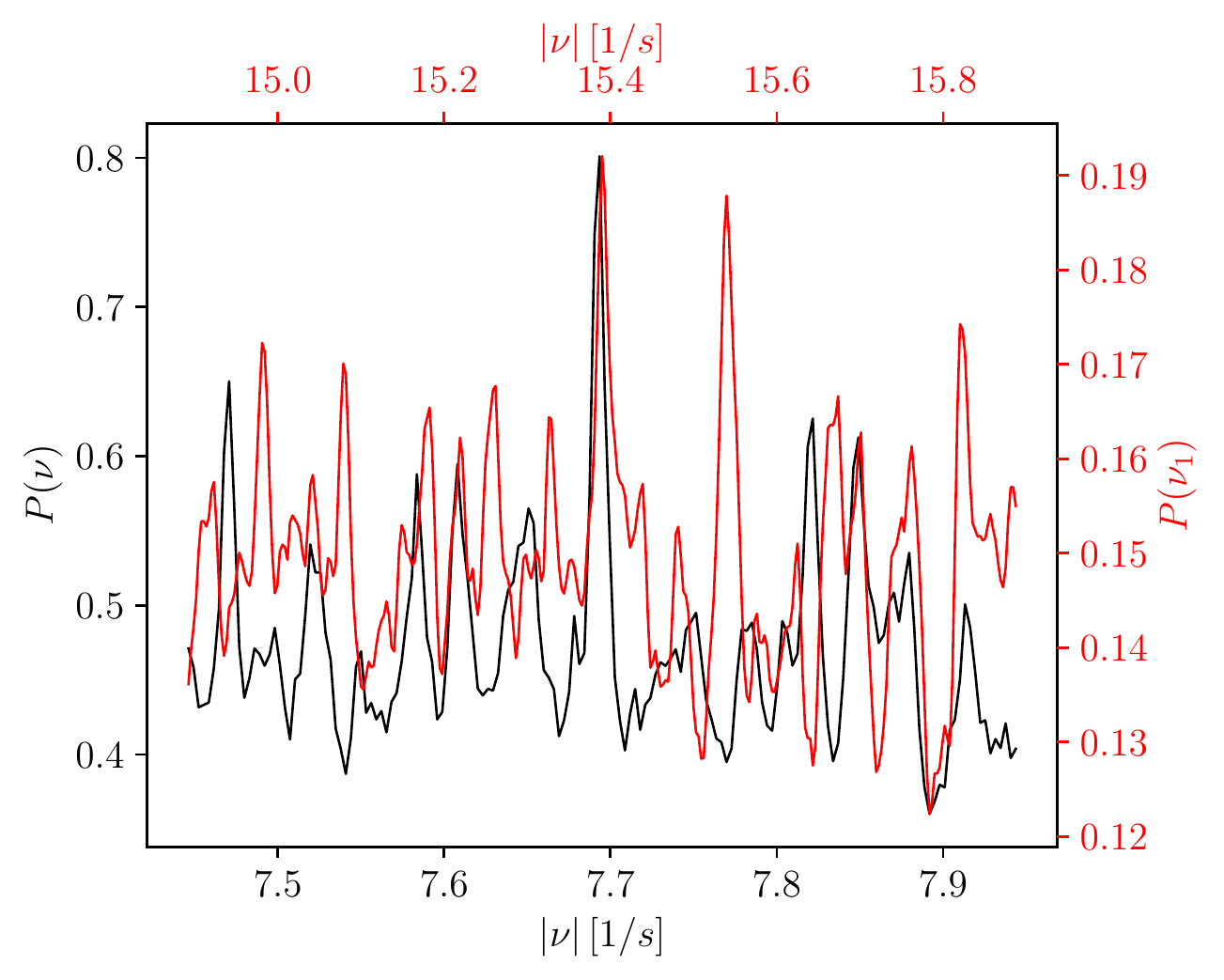}
\caption{Zoomed-in view of the reconstructed power spectra of the giant flare SGR 1900+14 around $\nu\sim7.693\,$Hz in black and its first overtone around $\nu\sim15.386\,$Hz in red. The plot corresponds to $\sigma_{\mathrm{sm}} = 1 \times 10^5$ as in \cref{sub_fig:SGR_spec_1e5}.}
\label{fig:zoom_1900_power}
\end{figure}

\begin{figure}[]
\centering
                        \includegraphics[width=.5\textwidth]{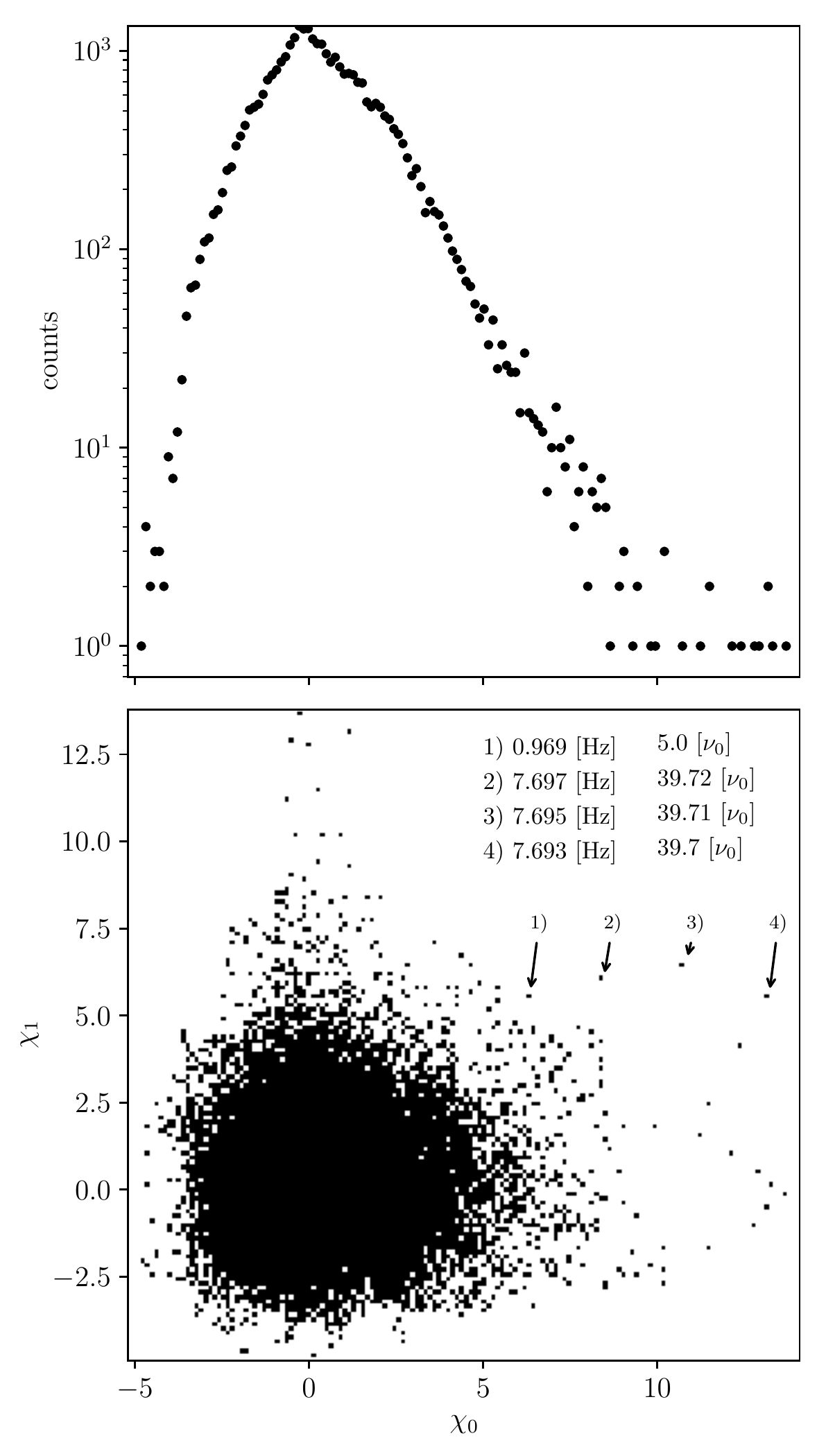}
\caption{Same as in \Cref{fig:hists_RXTE} for the 
SGR 1900+14 event, with $\sigma=  10^5$.
We indicate for all counts with $\chi_0 \geq 5$ and $\chi_1 
\geq 5$ the corresponding frequency in Hertz and its multiple of the neutron 
star frequency $\nu_0 = 0.1938\,$Hz.} 
\label{fig:hists_SGR}
\end{figure}

\begin{table}[]
\caption{All frequencies with $\chi_0 > 11$ and 
their multiplicity $n$ of the rotation period $\nu_0 = 0.1938$ [Hz] for SGR 
1900+14.}
\label{table:chi_SGR_10}
\centering
\begin{tabular}{c c c || c c c}
$\nu$ [Hz] & $\chi_0$ & $n$ [$\nu_0$]& $\nu$ [Hz] & $\chi_0$ & $n$ [$\nu_0$]\\
\hline
7.693 & 13.214 & 39.7 & 11.768 & 13.718 & 60.7 \\

\end{tabular}
\end{table}
With the same detection threshold as before, $\chi_0>11$, we find two candidate 
frequencies at $7.693\,$ and $11.768\,$Hz, see \Cref{table:chi_SGR_10}.
\begin{table}
\caption{Maximum $\chi_0$ at $\nu_\text{max}$ in a $5\%$ interval around 
previously observed oscillation frequencies $\nu$ for SGR 1900+14.
We also list the local variance $\sigma_\chi$ of $\chi_0$ at this interval.}
\label{table:chi_SGR}
\centering
\begin{tabular}{c c c c}
$\nu$ [Hz]  & $\nu_\text{max}\,$[Hz] in $\pm 5\%$ interval  & 
$\chi_{0,\text{max}}$ & 
$\sigma_{\chi}$ \\
\hline 
28.0 & 27.341 & 5.392  & 2.931\\
53.5 &  54.781 & 5.24 & 2.318\\
84 &  86.328 & 4.39 & 2.228\\
155.1& 161.719 & 3.777 & 1.582 
\end{tabular}
\end{table}
In \Cref{table:chi_SGR} we give the maximum $\chi_0$ around previously observed 
frequencies. As for SGR 1806-20, we cannot confirm any of the previously 
detected frequencies, the highest significance we see is $\chi_0>5.4$ for $28.0\,$ Hz.
We further investigated our combined criterium, but now reduced to
the giant flare of SGR 1900+14 compared to the flare of SGR 1806-20. \Cref{fig:hists_SGR} shows only one additional frequency at $\nu=7.695\,$Hz, 
which strengthens our confidence in the previous finding in \Cref{table:chi_SGR_10} at 
$\nu=7.693\,$Hz. In \Cref{fig:zoom_1900_power} we overplot the 
reconstructed power spectra around $\nu=7.693\,$Hz (black line) and its first 
overtone (red line). These two peaks are the largest in the given frequency range.

\begin{figure*}
\centering
\includegraphics[width=.875\textwidth]{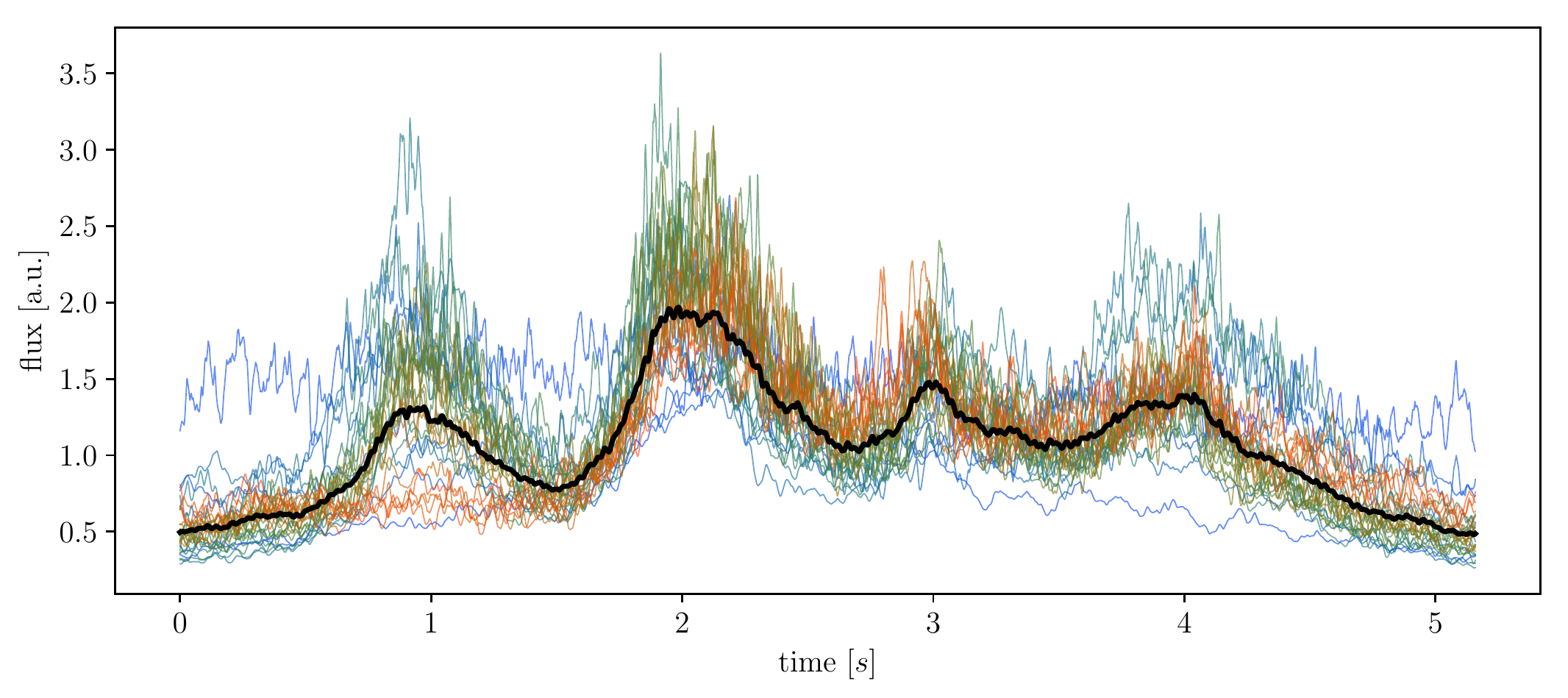}
\caption{Pulse profiles of different rotation periods 
overplotted for SGR 1900+14. The temporal evolution of 
the pulse profile is visible from blue to green to red.
Smooth curves without significant short time-variability are reconstructed 
during instrument deadtimes.
}
\label{fig:mean_period_SGR}
\end{figure*}

We plot the time evolution of the different pulses of SGR 1900+14 
in \Cref{fig:mean_period_SGR}. As before, blue indicates the beginning of the 
flare, green the middle, and red the end. Here, the data are of inferior 
quality compared to the SGR 1806-20 and the parts of the curves without 
significant short time-variability are purely reconstructed by our algorithm. 
For SGR 1900+14 we observed four maxima, which also evolved differently compared to 
each other. For example, the weakest maximum ($\mathrm{time}\sim1\,$s) declines almost 
monotonically with time, while the others remain rather constant  
with far less variability in time. \\ \\

In order to facilitate secondary studies based on our reconstructions and spectra, we provide
these data online at \url{ http://cdsweb.u-strasbg.fr/cgi-bin/qcat?J/A+A/}. At  \url{www.mpa-garching.mpg.de/ift/data/QPO} we also provides interactive plots of \Cref{fig:SGR_reconstruction,fig:SGR1806_reconstruction,fig:SGR1806_power}
for more detailed views. 

\section{Discussion}\label{sec:discussion}
We recovered the rotation period of the two magnetars $\nu^{1806}=0.1323\,$
Hz and $\nu^{1900}=0.1938\,$Hz, and up to the
$31$th overtone for SGR 1806-20 ($\chi_0>11$) and to the $18$th for SGR 
1900+14 ($\chi_0>7.5$). In addition, we only found potential periodic 
signals at $9.2\,$Hz (SGR1806-20) and $7.7\,$Hz (SGR1900+14), which are 
significant in $\chi_0$ and in the combination of $\chi_0$ and $\chi_1$ 
according to our respective criterium. For SGR 1900+14 and with the single 
criterium on $\chi_0$, we found another candidate frequency at $\nu=11.8\,$Hz, 
with a similar $\chi_0$ as the signal at $7.7\,Hz,$ but without detected 
overtones. Therefore we only consider $\nu=7.7\,$Hz as a potential signal.
The most robust feature is at $9.2\,$Hz for SGR 1806-20, it has the 
highest $\chi_0$ , and we detect some of its overtones as local maxima in 
$\chi_0$ at $18.3$ and $36.4\,$Hz, respectively. The two latter frequencies 
are similar to previously reported frequencies at $17.9$ \citep{Watts2006} and 
$36.8\,$Hz 
\citep{Hambaryan2011}.

Although the reimplemented Bayesian inference algorithm D$^3$PO proved (see \Cref{sec:performance,sec:injection,sec:QPO_injection}) its 
basic ability
to reconstruct QPO in photon bursts in many tests, even in the low S/N regime,
we cannot confirm the previously reported frequencies  at
$17, 21, 26, 29, 59, 92.5, 116,$ and $150\,$Hz for SGR 1806-20 and $28, 53, 84,$ and  
$155\,$Hz for SGR 1900+14. In particular, we show in 
\Cref{sec:QPO_injection} that our algorithm is able to recover 
sufficiently strong quasi-periodic signals around $90\,$Hz with a width of 
$\sim 0.6\,$Hz.

Our analysis and therefore all reconstructed power spectra and 
reconstructed photon fluxes depend on the particular chosen smoothness-enforcing prior $\sigma_{\mathrm{sm}}$ . Selecting a smaller 
$\sigma_{\mathrm{sm}}$ allows the algorithm to denoise the power spectrum even 
at low S/N. However, at high frequencies, the small $\sigma_{\mathrm{sm}}$ 
leads to a lower sensitivity for spectral lines. Therefore, a trade-off needs to be found for $\sigma_{\mathrm{sm}}$ between better denoising (small 
$\sigma_{\mathrm{sm}}$) and better sensitivity (large $\sigma_{\mathrm{sm}}$).
Owing to the sharply deteriorating S/N for frequency ranges around $625\,$Hz and 
$1840\,$Hz, which would require a very small $\sigma_{\mathrm{sm}}$, we are 
currently not able to investigate these frequencies for QPOs.

We assume that our findings hold and determine the effect that
the results have on the interpretation of 
the theoretical model of magnetar oscillations. 
First of all, the new candidate frequencies at $9.2\,$ and $7.7\,$Hz 
for SGR1806-20 and SGR1900+14, respectively, are much lower than the 
frequencies reported so far. Here, our method has a clear advantage over 
previous studies, as we can analyse the entire data set at once, meaning that no parts of the data set are left out. In principle, this leads to a more robust reconstruction and also counteracts statistical artefacts that may be introduced by a windowed analysis. 
Our method is fully noise-aware and has in principle no problems with observational 
dead times. 
If these two frequencies are related to neutron star oscillations, the 
parameters of current models change significantly. When we assume
that the oscillations 
are pure crustal shear modes, the identification of 
the low frequency with the $n=0, l=2$ mode would favour models with rather low 
shear speeds and therefore EoS with a fast increase of symmetry 
energy \citep{Steiner2009}.  However, pure shear models are not very likely because of the strong 
interaction with the magnetic field in the interior of the star 
\citep{Levin2006,Levin2007,Gabler2011letter,Gabler2012}. If the magnetic field is not neglected, coupled magneto-elastic oscillations
need to be considered. In this case, the much lower 
fundamental oscillation frequency indicates a magnetic field of the order 
of $\bar B\sim 6\times10^{13} - 3\times10^{14}\,$G, lower by a factor of $\sim 3$ than our estimates in 
\cite{Gabler2013b}. With such weak magnetic fields, the oscillations 
would also remain confined to the core for models with very strong shear modulus, 
and there would be no chance to observe the oscillations 
exterior to the star. 
Therefore, similarly to the case of pure crustal shear oscillations, 
the low frequencies of $9.2\,$ or $7.7\,$Hz favour EoSs that give a small shear 
modulus. 
However, the estimation of the magnetic field strength has some serious 
problems : i) The spin-down estimate is only accurate to a factor of 
a few. ii) The particular magnetic field configuration inside a magnetar 
is unknown, meaning that even for dipolar-like fields in the exterior, there are 
different possible realization in the interior. iii) In addition
to the degeneracies 
of  the dipolar magnetic field strength and the magnetic field configuration, which lead to comparable frequencies, the compactness of the neutron star 
and the superfluid properties of the core matter also influence the frequencies 
significantly \citep{Gabler2017}. In order to advance and to lift some of these 
degeneracies, we need more observations of QPOs, and if possible,
observations for different 
sources.

In respect of the potential of improving our method by modelling the spectra as an independent combination 
of a continuum and lines, we postpone a more thorough 
discussion of these consequences to future work. We also plan to improve our 
algorithm by taking the deadtime after each photon detection into account and 
include further instrument responses. However, we expect these improvements to 
increase our detection limits only by about a few per cent.

We have furthermore shown the capability of our code D$^3$PO to denoise and 
recover the shape of the light curve. This may have interesting applications 
in the field of modelling the pulses of X-ray bursters.

\section{Conclusion}\label{sec:conclusions}
We have applied a new Bayesian method, D$^3$PO to analyse time-series data of the giant flares of SGR 1806-20 and SGR 1900+14. 
Thereby, we ensured that the photon shot noise, operational down times of the instrument, and the 
\textit{\textup{a priori}} unknown power spectrum of the photon flux were taken into account. In order to denoise and reconstruct the logarithmic photon flux and its power spectrum simultaneously from the data, we had to assume some kind of spectral smoothness on the power spectrum to counteract data variance. 
Our analysis cannot confirm previous findings of QPOs in the giant flare data, 
but we found new candidates for periodic signals at $9.2\,$Hz for SGR 
1806-20 and $7.7\,$Hz for SGR1900+14. If these are real and related to the lowest 
frequency oscillation of the magnetar, our results favour high-compactness,
weaker magnetic fields than were assumed before, and low shear moduli. 
The necessary application of a smoothness-enforcing prior results in a reduced sensitivity to the spectral line in the power spectrum. Hence, we propose for future work to decompose the power spectrum into two components, one smooth background spectrum, and one consisting of spectral lines. In this way, the previously reported lines may be confirmed or refuted with higher confidence, or additional and as yet unknown frequencies in QPOs might be
discovered. 

\begin{acknowledgements}
We would like to thank Martin Reinecke for his computational and numerical support, and Ewald M{\"u}ller and the anonymous referee for their fruitful discussions and comments. Furthermore, we would like to acknowledge Anna Watts, Phil Uttley, and Daniela Huppenkothen for a controversial debate. 
Work supported by the ERC Advanced Grant
no. 341157-COCO2CASA.
This research has made use of data and/or software provided by the High Energy 
Astrophysics Science Archive Research Center (HEASARC), which is a service of 
the Astrophysics Science Division at NASA/GSFC and the High Energy Astrophysics 
Division of the Smithsonian Astrophysical Observatory.

\end{acknowledgements}

\begin{appendix}
\section{Influence of the smoothness-enforcing parameter $\sigma$}
\label{sec:performance}
\begin{figure}
\centering
        \begin{subfigure}{\hsize}
        \centering
                        \includegraphics[width=.875\textwidth]{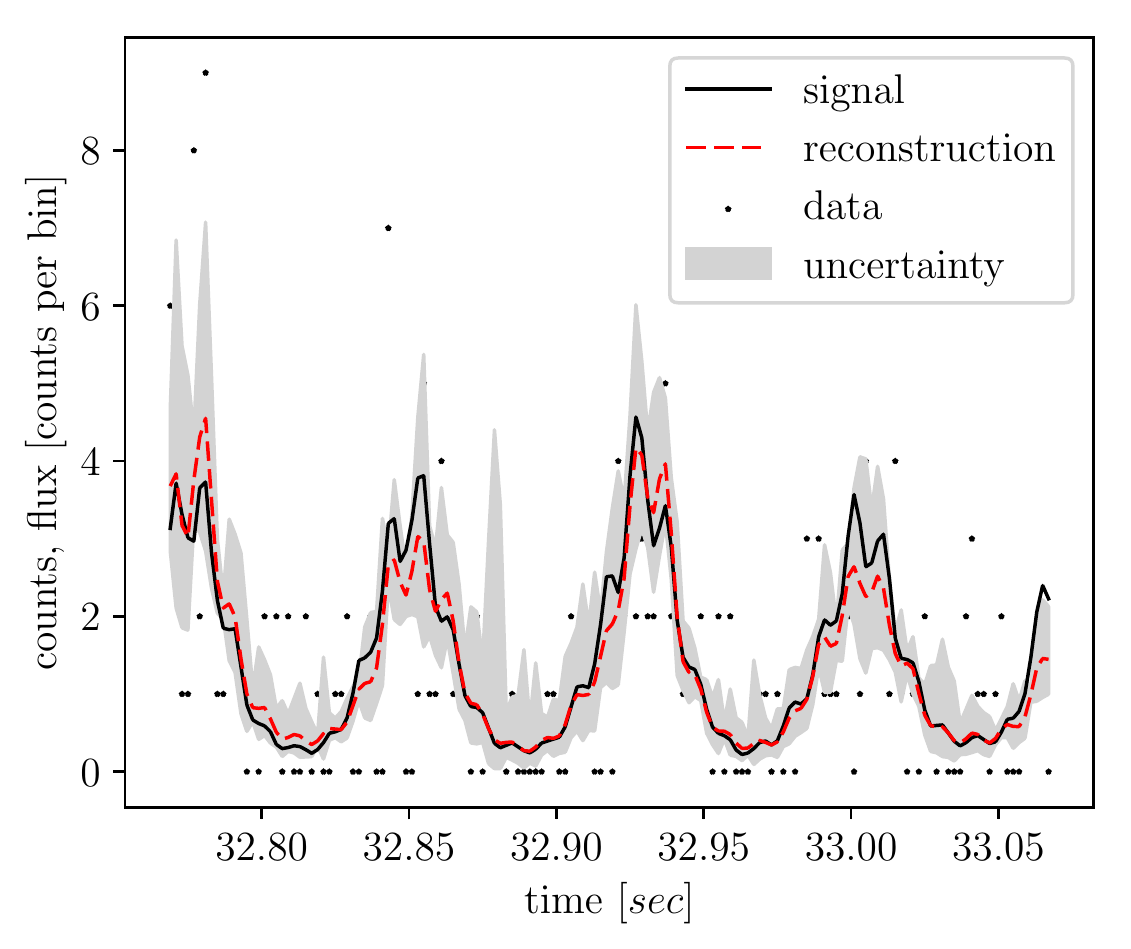}
                        \caption{}
                        \label{sub_fig:tests_snapshot}
        \end{subfigure}
        \begin{subfigure}{\hsize}
        \centering
                        \includegraphics[width=.875\textwidth]{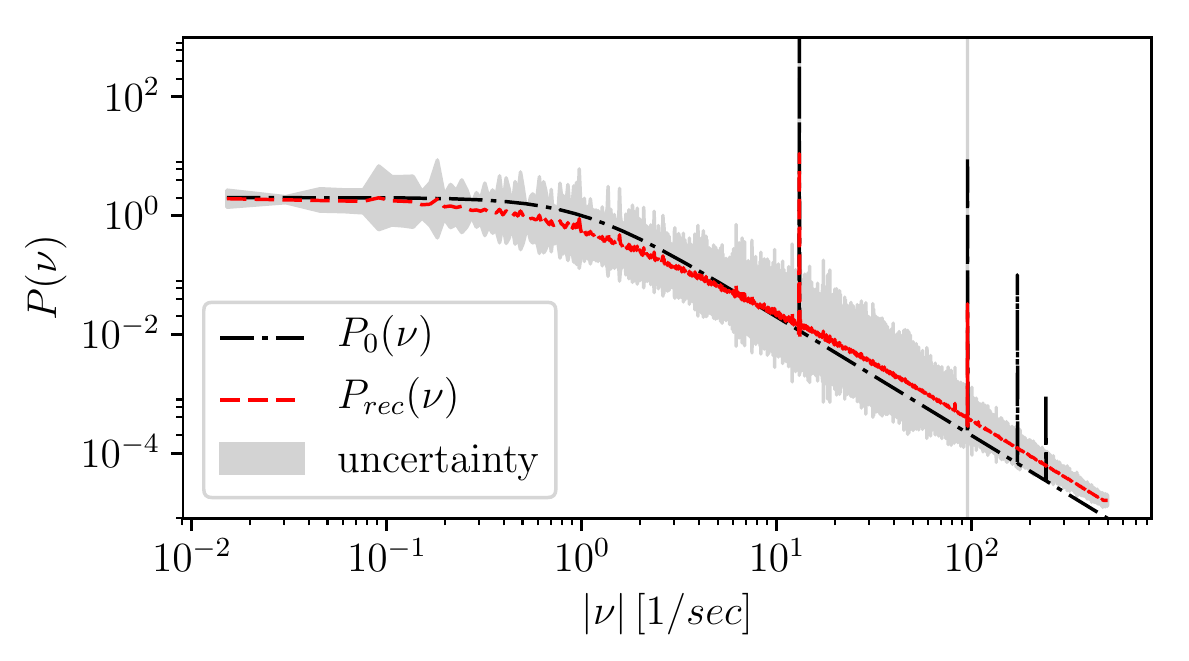}
                        \caption{}
                        \label{sub_fig:tests_spec_5e5}
        \end{subfigure}
        \begin{subfigure}{\hsize}
        \centering
                        \includegraphics[width=.875\textwidth]{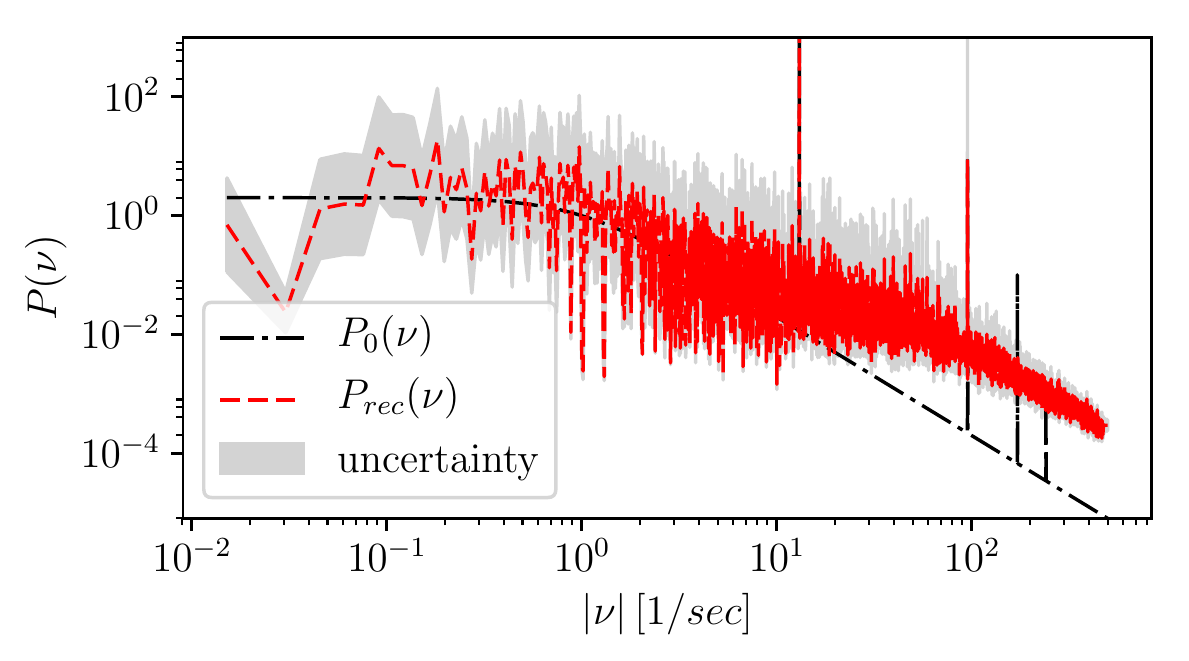}
                        \caption{}
                        \label{sub_fig:tests_spec_1e5}
        \end{subfigure}
    \caption{Reconstruction, i.e. map and power spectra from mock data, created according to \Cref{sec:performance}. For clarity, \Cref{sub_fig:tests_snapshot} only shows a snapshot of the events between $32.8 \, \text{sec}\lesssim t \lesssim 33 \, \text{sec}$. In addition to the raw photon counts, the black line shows the original signal $\vec{s}$, as well as the reconstruction in red, including its uncertainty. \Cref{sub_fig:tests_spec_5e5} shows the reconstructed power spectra as well as its uncertainty, as
in \Cref{sub_fig:tests_snapshot}, using a smoothness-enforcing prior with $\sigma_{\mathrm{sm}} = 10^{4}$. \Cref{sub_fig:tests_spec_1e5} shows a reconstructed power spectrum with a weaker smoothness-enforcing prior with $\sigma_{\mathrm{sm}}=10^{5}$.}
\label{fig:algo_tests}
\end{figure}

To demonstrate the performance of the Bayesian inference algorithm, we challenged the new implementation of the algorithm with mock data. To do so, we drew Poisson samples from $\lambda = \vec{R} e^{\vec{s}}$. For simplicity, we assumed $\vec{R}$ to be the identity operator. The signal field $\vec{s}$ was drawn from a Gaussian random field $\vec{s} \hookleftarrow \mathscr{G}(s, S)$, with 
\begin{align}
S_{\nu \nu'} = \frac{27}{\left(1+\nu/\nu_0 \right)^2} \delta_{\nu \nu'}\,,
\end{align}
with $\nu_0 = 1$ absorbing the physical units. To test whether the algorithm can reconstruct discrete frequencies as well, we further injected four $\delta$-peaks into $S_{\nu \nu'}$ at randomly chosen positions. The peak heights decrease at higher frequencies to be as realistic as possible. For the test we used a one-dimensional regular grid with with $2^{16}$ pixels, each with a volume of $1/1000$ seconds. \\
In \Cref{fig:algo_tests} we tested the principle capabilities of the algorithm under the above-mentioned setting while varying the strength of the smoothness enforcing parameter $\sigma_{\mathrm{sm}}$. In the scenario of $\sigma_{\mathrm{sm}} =  10^{4}$, shown in \Cref{sub_fig:tests_snapshot,sub_fig:tests_spec_5e5}, the algorithm was able to recover the signal as well as the power spectrum well within the one-$\sigma$ confidence interval. However, the two highest injected frequencies could not be recovered, as their injected strength and therefore their effect on the data was too weak to be distinguished from the Poisson shot noise. \\
In comparison to \Cref{sub_fig:tests_spec_5e5}, \Cref{sub_fig:tests_spec_1e5} shows the reconstructed power spectrum using a significantly weaker smoothness enforcing $\sigma_{\mathrm{sm}} =  10^{5}$. As a first obvious consequence, the local variance of the spectrum increases by multiple orders of magnitude. However, the same two frequencies as with the stronger $\sigma_{\mathrm{sm}}$ were clearly recovered. As a further consequence of the smaller $\sigma_{\mathrm{sm}}$ , the power spectrum does not fully recover the shape of $S_{\nu \nu'}$ as it picks up more noise features from the recovered map and data. For a detailed discussion and mathematical derivation of this smoothness-enforcing parameter, we refer to \cite{2013PhRvE..87c2136O}. \\ 

\section{Recovery of spectral lines at high frequencies}
\label{sec:injection}
\begin{figure}
\centering
        \begin{subfigure}{\hsize}
        \centering
                        \includegraphics[width=.875\textwidth]{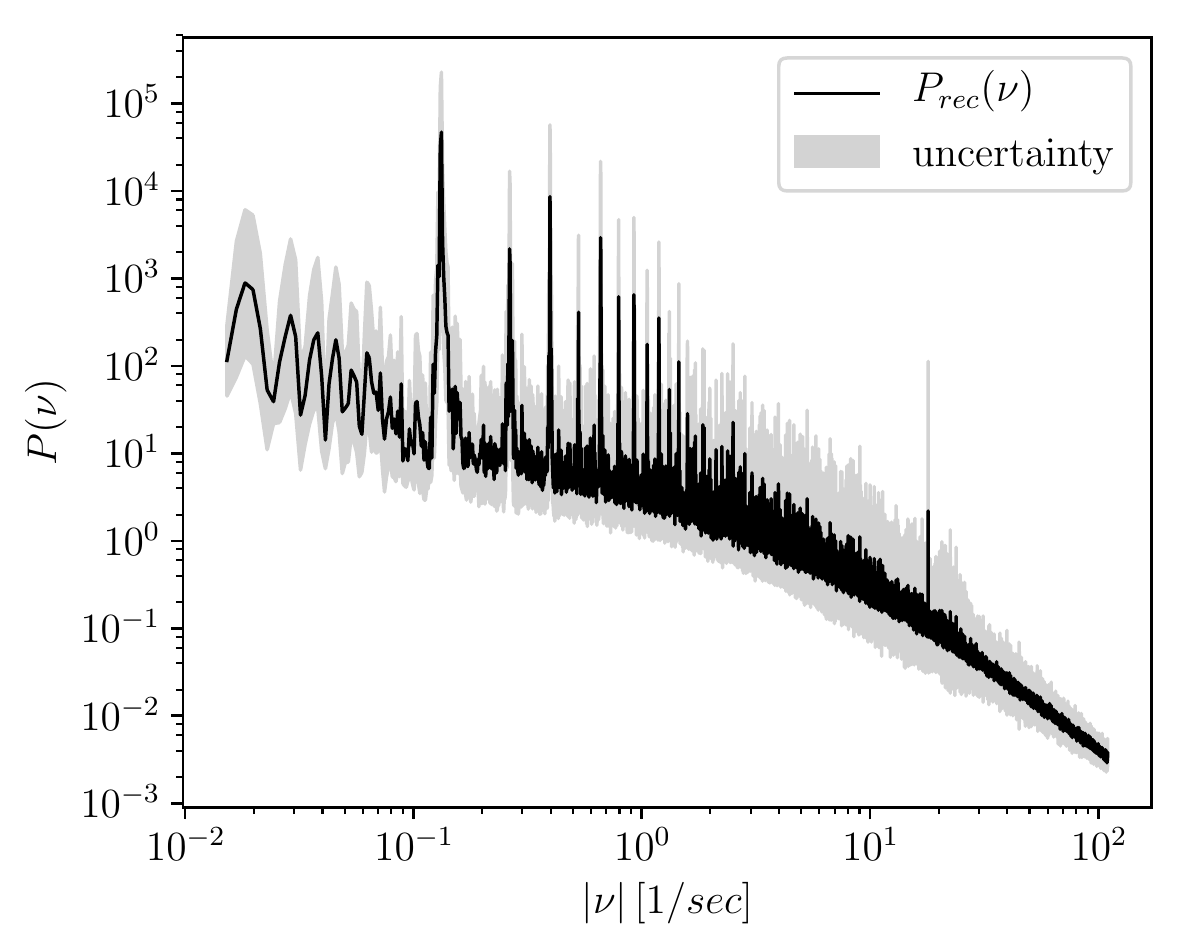}
                        \caption{}
                        \label{sub_fig:injection_1e2}
        \end{subfigure}
        \begin{subfigure}{\hsize}
        \centering
                        \includegraphics[width=.875\textwidth]{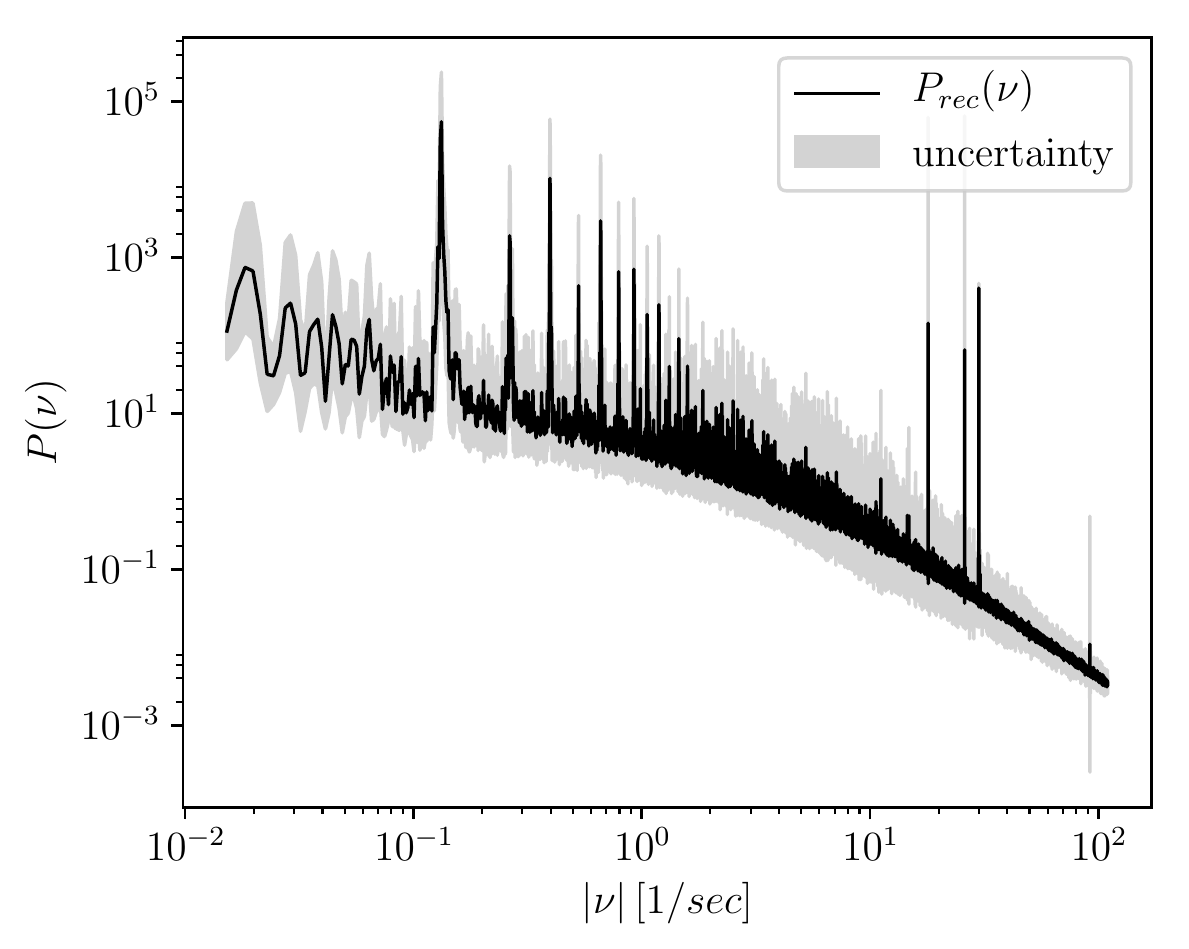}
                        \caption{}
                        \label{sub_fig:injection_1e5}
        \end{subfigure}
    \caption{Reconstructed power spectrum, according to the test scenario described in \Cref{sec:rec_lines}. The strengths of the injected frequencies are $10^2$ (top) and $10^5$ (bottom) times stronger than the local power at these frequencies. The displayed uncertainties indicate the one-$\sigma$ confidence interval.}
\label{fig:injection}
\end{figure}

Now we estimate how strong periodic oscillations must be in 
order to be detectable in the reconstructed power spectrum. To be as close as 
possible to a realistic scenario, we manipulated the reconstructed photon flux of 
event SGR 1806-20. We added a periodic signal with discrete frequencies at 
$18.0, 26.0, 30.0,$ and $90.0$ Hz to the reconstructed $\vec{\phi}$ shown in 
\Cref{sub_fig:RXTE_whole}. The relative strength of this additional photon flux 
was varied, between $10^2$ and $10^5$ times stronger than the local power 
of the injected frequencies. From these manipulated fluxes, we
again drew Poisson 
samples and let the algorithm recover the power spectrum and the flux. \\
\Cref{fig:injection} shows the reconstructed power spectra. In the setting of the top panel, we were only able to recover the injected frequency at $18$ Hz, the other three could not be inferred as their strength of $10^2$ compared to the local power was too weak to be identified as part of the signal and not the Poisson shot noise. However, as the one-$\sigma$ confidence interval of the reconstructed spectrum at $18 \, \text{Hz} \lesssim \nu  \lesssim 40 \, \text{Hz} $ increases, the algorithm is just at the S/N threshold at which it is able to reconstruct discrete frequencies. If the injected strength of the frequencies, that
is, the amplitude of the injected signal, becomes larger, as in \Cref{sub_fig:injection_1e5}, the algorithm can infer all frequencies. \\
Hence, we may conclude that if the data show periodic signals at high frequencies with sufficient strength, they can be reconstructed by the algorithm. The strength of the periodic signal must therefore be at least $10^2$ times stronger than the local power of the signal in order to give a significant imprint on the reconstructed power spectrum. 

\label{sec:rec_lines}

\section{Recovery of quasi-periodic oscillations}
\label{sec:QPO_injection}
\begin{figure}
\includegraphics[width=.5\textwidth]{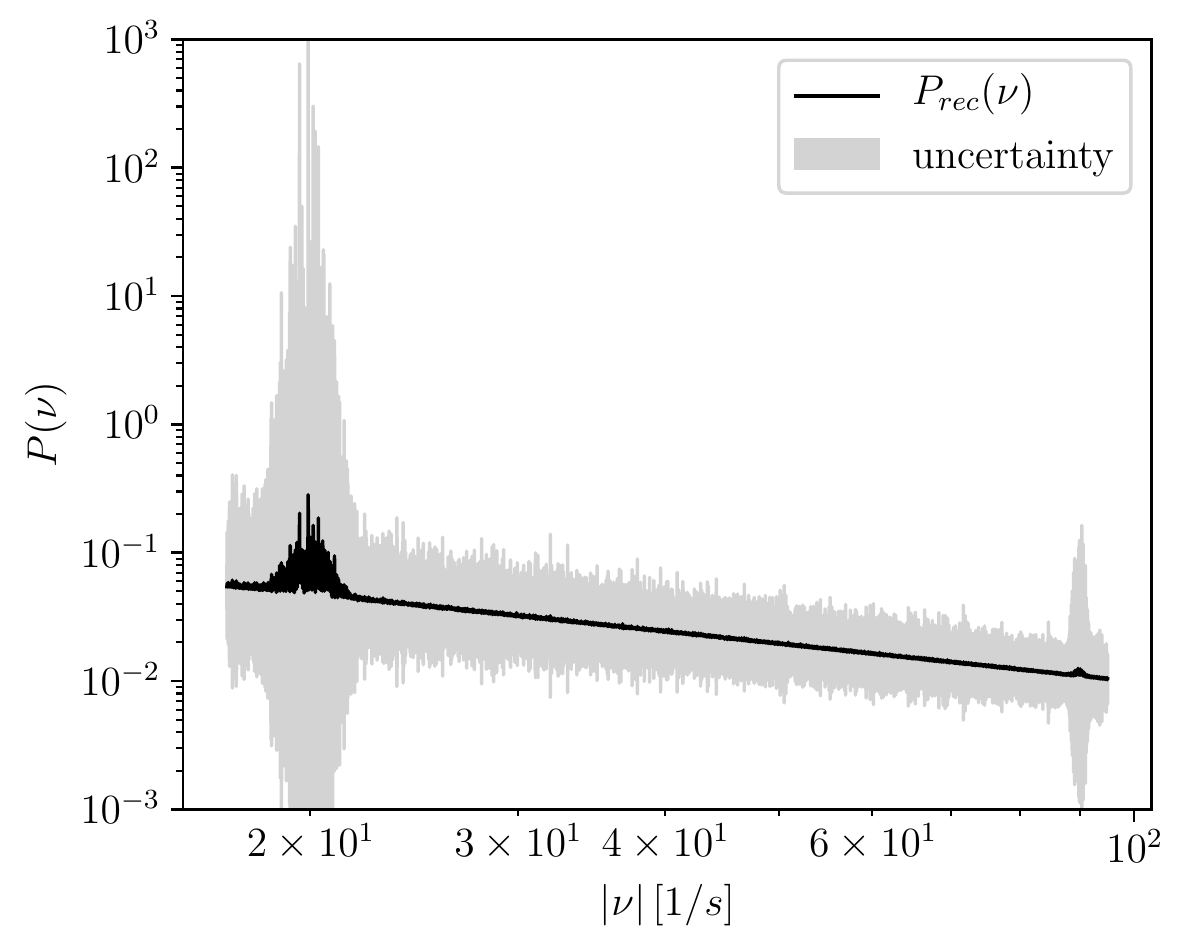}
\caption{Reconstructed power spectrum according to the QPO injection
test described in \Cref{sec:QPO_injection}. The displayed uncertainties indicate 
the one-$\sigma$ confidence interval.} 
\label{fig:QPO_injection}
\end{figure}

Here, we estimate how strong QPOs have to be for our algorithm to still find 
a significant signal in the reconstructed power spectrum. Similar to 
\Cref{sec:injection}, we used the reconstructed photon flux, 
\Cref{sub_fig:RXTE_whole} of event SGR 1806-20, and added a quasi-periodic signal 
at around $20$ Hz and $90$ Hz. We assumed that the power spectrum of a QPO 
has an approximately Gaussian shape with variance $\sigma \approx 0.6$. 
From this manipulated flux, we drew Poisson samples and let the algorithm 
recover the power spectrum and the flux.
\begin{figure}
\includegraphics[width=.5\textwidth]{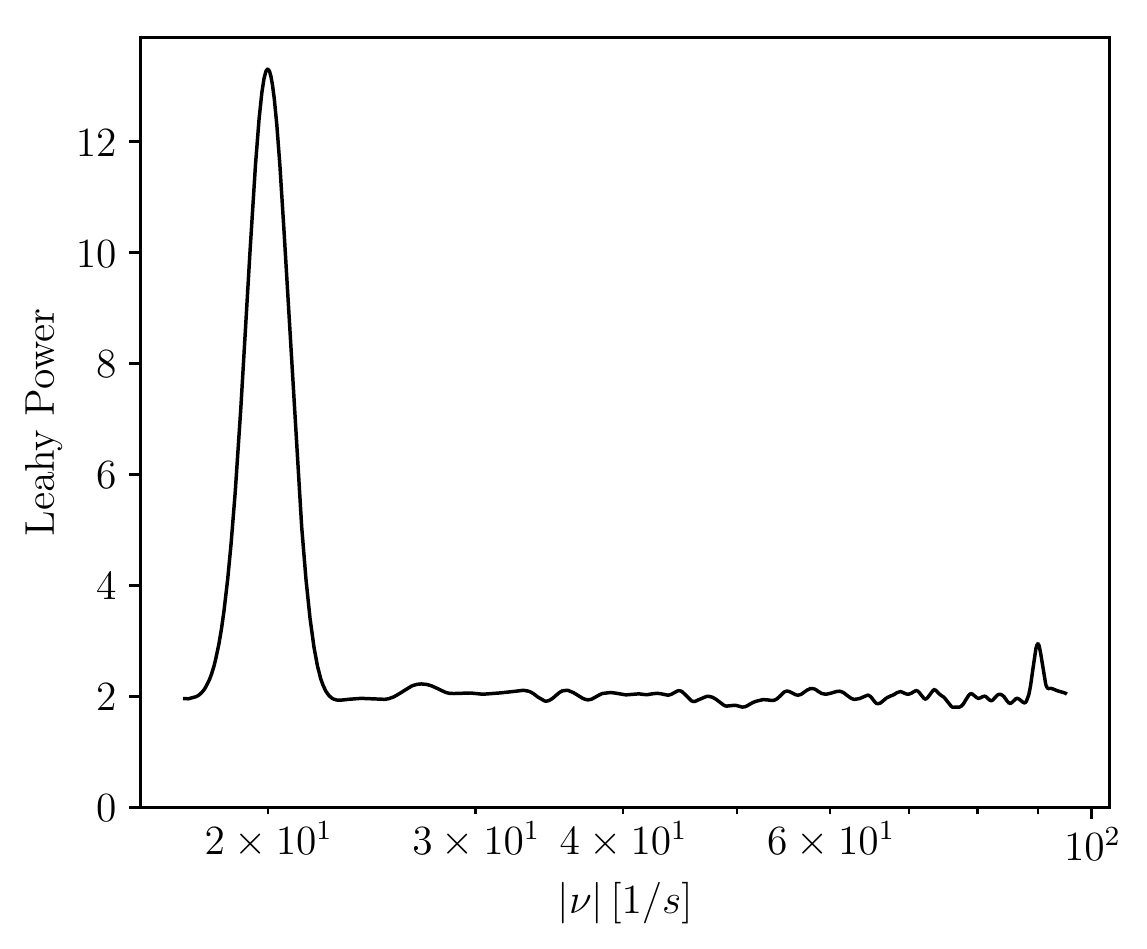}
\caption{Smoothed Leahy power of the injected pure QPO-like signal. The two QPOs at $20$ Hz 
and $90$ Hz were reconstructed by D$^3$PO, \Cref{fig:QPO_injection}.}
\label{fig:leahy}
\end{figure}
\Cref{fig:leahy} shows the power spectrum of the Poissonian photon counts for a pure QPO-like injected signal, 
smoothed with a Gaussian convolution kernel whose variance is the same as that of the injected QPO. The reconstructed power spectrum is shown in  
\Cref{fig:QPO_injection}. The S/N around $20$ Hz is much higher than 
the S/N around $90$ Hz. Therefore, the QPO at $20$ Hz has a significantly 
larger amplitude in the reconstructed power spectrum than that around 
$90$ Hz. The strength of the injected QPO at $90$ Hz is just at the 
threshold to give a significant signal in the reconstructed power spectrum. 
Thus, we may conclude that at this frequency, a QPO has to have a Leahy power 
of the order of $3$ in order be detectable by D$^3$PO. 

\end{appendix}

\bibliographystyle{aa}
\bibliography{bibliography}

\begin{thebibliography}{45}
\expandafter\ifx\csname natexlab\endcsname\relax\def\natexlab#1{#1}\fi

\bibitem[{{Cerd{\'a}-Dur{\'a}n} {et~al.}(2009){Cerd{\'a}-Dur{\'a}n},
  {Stergioulas}, \& {Font}}]{Cerda2009}
{Cerd{\'a}-Dur{\'a}n}, P., {Stergioulas}, N., \& {Font}, J.~A. 2009, \mnras,
  397, 1607

\bibitem[{{Colaiuda} {et~al.}(2009){Colaiuda}, {Beyer}, \&
  {Kokkotas}}]{Colaiuda2009}
{Colaiuda}, A., {Beyer}, H., \& {Kokkotas}, K.~D. 2009, \mnras, 396, 1441

\bibitem[{{Colaiuda} \& {Kokkotas}(2011)}]{Colaiuda2011}
{Colaiuda}, A. \& {Kokkotas}, K.~D. 2011, \mnras, 414, 3014

\bibitem[{{Deibel} {et~al.}(2014){Deibel}, {Steiner}, \& {Brown}}]{Deibel2014}
{Deibel}, A.~T., {Steiner}, A.~W., \& {Brown}, E.~F. 2014, \prc, 90, 025802

\bibitem[{{Duncan}(1998)}]{Duncan1998}
{Duncan}, R.~C. 1998, \apjl, 498, L45

\bibitem[{{En{\ss}lin} \& {Frommert}(2011)}]{2011PhRvD..83j5014E}
{En{\ss}lin}, T.~A. \& {Frommert}, M. 2011, \prd, 83, 105014

\bibitem[{{En{\ss}lin} \& {Knollm{\"u}ller}(2016)}]{2016arXiv161208406E}
{En{\ss}lin}, T.~A. \& {Knollm{\"u}ller}, J. 2016, ArXiv e-prints
  [\eprint[arXiv]{1612.08406}]

\bibitem[{{Gabler} {et~al.}(2011){Gabler}, {Cerd{\'a} Dur{\'a}n}, {Font},
  {M{\"u}ller}, \& {Stergioulas}}]{Gabler2011letter}
{Gabler}, M., {Cerd{\'a} Dur{\'a}n}, P., {Font}, J.~A., {M{\"u}ller}, E., \&
  {Stergioulas}, N. 2011, \mnras, 410, L37

\bibitem[{{Gabler} {et~al.}(2012){Gabler}, {Cerd{\'a}-Dur{\'a}n},
  {Stergioulas}, {Font}, \& {M{\"u}ller}}]{Gabler2012}
{Gabler}, M., {Cerd{\'a}-Dur{\'a}n}, P., {Stergioulas}, N., {Font}, J.~A., \&
  {M{\"u}ller}, E. 2012, \mnras, 421, 2054

\bibitem[{{Gabler} {et~al.}(2013){Gabler}, {Cerd{\'a}-Dur{\'a}n},
  {Stergioulas}, {Font}, \& {M{\"u}ller}}]{Gabler2013b}
{Gabler}, M., {Cerd{\'a}-Dur{\'a}n}, P., {Stergioulas}, N., {Font}, J.~A., \&
  {M{\"u}ller}, E. 2013, Physical Review Letters, 111, 211102

\bibitem[{{Gabler} {et~al.}(2016){Gabler}, {Cerd{\'a}-Dur{\'a}n},
  {Stergioulas}, {Font}, \& {M{\"u}ller}}]{Gabler2016}
{Gabler}, M., {Cerd{\'a}-Dur{\'a}n}, P., {Stergioulas}, N., {Font}, J.~A., \&
  {M{\"u}ller}, E. 2016, \mnras, 460, 4242

\bibitem[{{Gabler} {et~al.}(2017){Gabler}, {Cerd{\'a}-Dur{\'a}n},
  {Stergioulas}, {Font}, \& {M{\"u}ller}}]{Gabler2017}
{Gabler}, M., {Cerd{\'a}-Dur{\'a}n}, P., {Stergioulas}, N., {Font}, J.~A., \&
  {M{\"u}ller}, E. 2017, in preparation

\bibitem[{{Glampedakis} {et~al.}(2011){Glampedakis}, {Andersson}, \&
  {Samuelsson}}]{Glampedakis2011a}
{Glampedakis}, K., {Andersson}, N., \& {Samuelsson}, L. 2011, \mnras, 410, 805

\bibitem[{{Glampedakis} {et~al.}(2006){Glampedakis}, {Samuelsson}, \&
  {Andersson}}]{Glampedakis2006b}
{Glampedakis}, K., {Samuelsson}, L., \& {Andersson}, N. 2006, \mnras, 371, L74

\bibitem[{{Hambaryan} {et~al.}(2011){Hambaryan}, {Neuh{\"a}user}, \&
  {Kokkotas}}]{Hambaryan2011}
{Hambaryan}, V., {Neuh{\"a}user}, R., \& {Kokkotas}, K.~D. 2011, \aap, 528,
  A45+

\bibitem[{{Huppenkothen} {et~al.}(2014{\natexlab{a}}){Huppenkothen},
  {D'Angelo}, {Watts}, {Heil}, {van der Klis}, {van der Horst}, {Kouveliotou},
  {Baring}, {G{\"o}{\u g}{\"u}{\c s}}, {Granot}, {Kaneko}, {Lin}, {von
  Kienlin}, \& {Younes}}]{Huppenkothen2014a}
{Huppenkothen}, D., {D'Angelo}, C., {Watts}, A.~L., {et~al.}
  2014{\natexlab{a}}, \apj, 787, 128

\bibitem[{{Huppenkothen} {et~al.}(2014{\natexlab{b}}){Huppenkothen}, {Heil},
  {Watts}, \& {G{\"o}{\u g}{\"u}{\c s}}}]{Huppenkothen2014b}
{Huppenkothen}, D., {Heil}, L.~M., {Watts}, A.~L., \& {G{\"o}{\u g}{\"u}{\c
  s}}, E. 2014{\natexlab{b}}, \apj, 795, 114

\bibitem[{{Huppenkothen} {et~al.}(2014{\natexlab{c}}){Huppenkothen}, {Watts},
  \& {Levin}}]{Huppenkothen2014c}
{Huppenkothen}, D., {Watts}, A.~L., \& {Levin}, Y. 2014{\natexlab{c}}, \apj,
  793, 129

\bibitem[{{Israel} {et~al.}(2005){Israel}, {Belloni}, {Stella}, {Rephaeli},
  {Gruber}, {Casella}, {Dall'Osso}, {Rea}, {Persic}, \&
  {Rothschild}}]{Israel2005}
{Israel}, G.~L., {Belloni}, T., {Stella}, L., {et~al.} 2005, \apjl, 628, L53

\bibitem[{{Junklewitz} {et~al.}(2016){Junklewitz}, {Bell}, {Selig}, \&
  {En{\ss}lin}}]{2016A&amp;A...586A..76J}
{Junklewitz}, H., {Bell}, M.~R., {Selig}, M., \& {En{\ss}lin}, T.~A. 2016,
  \aap, 586, A76

\bibitem[{{Levin}(2006)}]{Levin2006}
{Levin}, Y. 2006, \mnras, 368, L35

\bibitem[{{Levin}(2007)}]{Levin2007}
{Levin}, Y. 2007, \mnras, 377, 159

\bibitem[{{Messios} {et~al.}(2001){Messios}, {Papadopoulos}, \&
  {Stergioulas}}]{Messios2001}
{Messios}, N., {Papadopoulos}, D.~B., \& {Stergioulas}, N. 2001, \mnras, 328,
  1161

\bibitem[{{Olausen} \& {Kaspi}(2014)}]{2014ApJS..212....6O}
{Olausen}, S.~A. \& {Kaspi}, V.~M. 2014, \apjs, 212, 6

\bibitem[{{Oppermann} {et~al.}(2013){Oppermann}, {Selig}, {Bell}, \&
  {En{\ss}lin}}]{2013PhRvE..87c2136O}
{Oppermann}, N., {Selig}, M., {Bell}, M.~R., \& {En{\ss}lin}, T.~A. 2013, \pre,
  87, 032136

\bibitem[{{Passamonti} \& {Lander}(2013)}]{Passamonti2013}
{Passamonti}, A. \& {Lander}, S.~K. 2013, \mnras, 429, 767

\bibitem[{{Passamonti} \& {Lander}(2014)}]{Passamonti2014}
{Passamonti}, A. \& {Lander}, S.~K. 2014, \mnras, 438, 156

\bibitem[{{Piro}(2005)}]{Piro2005}
{Piro}, A.~L. 2005, \apjl, 634, L153

\bibitem[{{Pumpe} {et~al.}(2016){Pumpe}, {Greiner}, {M{\"u}ller}, \&
  {En{\ss}lin}}]{2016PhRvE..94a2132P}
{Pumpe}, D., {Greiner}, M., {M{\"u}ller}, E., \& {En{\ss}lin}, T.~A. 2016,
  \pre, 94, 012132

\bibitem[{{Samuelsson} \& {Andersson}(2007)}]{Samuelsson2007}
{Samuelsson}, L. \& {Andersson}, N. 2007, \mnras, 374, 256

\bibitem[{{Samuelsson} \& {Andersson}(2009)}]{Samuelsson2009}
{Samuelsson}, L. \& {Andersson}, N. 2009, Classical and Quantum Gravity, 26,
  155016

\bibitem[{Selig \& En{\ss}lin(2015)}]{Selig:2015rt}
Selig, M. \& En{\ss}lin, T. 2015, A\&A, 574, A74

\bibitem[{Selig {et~al.}(2015)Selig, Vacca, Oppermann, \&
  En{\ss}lin}]{Selig:2015ul}
Selig, M., Vacca, V., Oppermann, N., \& En{\ss}lin, T.~A. 2015, A\&A, 581, A126

\bibitem[{{Sotani} {et~al.}(2016){Sotani}, {Iida}, \& {Oyamatsu}}]{Sotani2016}
{Sotani}, H., {Iida}, K., \& {Oyamatsu}, K. 2016, \na, 43, 80

\bibitem[{{Sotani} {et~al.}(2007){Sotani}, {Kokkotas}, \&
  {Stergioulas}}]{Sotani2007}
{Sotani}, H., {Kokkotas}, K.~D., \& {Stergioulas}, N. 2007, \mnras, 375, 261

\bibitem[{{Sotani} {et~al.}(2008){Sotani}, {Kokkotas}, \&
  {Stergioulas}}]{Sotani2008}
{Sotani}, H., {Kokkotas}, K.~D., \& {Stergioulas}, N. 2008, \mnras, 385, L5

\bibitem[{{Sotani} {et~al.}(2013){Sotani}, {Nakazato}, {Iida}, \&
  {Oyamatsu}}]{Sotani2013}
{Sotani}, H., {Nakazato}, K., {Iida}, K., \& {Oyamatsu}, K. 2013, \mnras, 428,
  L21

\bibitem[{{Steiner} \& {Watts}(2009)}]{Steiner2009}
{Steiner}, A.~W. \& {Watts}, A.~L. 2009, Physical Review Letters, 103, 181101

\bibitem[{{Steininger} {et~al.}(2017){Steininger}, {Dixit}, {Frank}, {Greiner},
  {Hutschenreuter}, {Knollm{\"u}ller}, {Leike}, {Porqueres}, {Pumpe},
  {Reinecke}, {{\v S}raml}, {Varady}, \& {En{\ss}lin}}]{2017arXiv170801073S}
{Steininger}, T., {Dixit}, J., {Frank}, P., {et~al.} 2017, ArXiv e-prints
  [\eprint[arXiv]{1708.01073}]

\bibitem[{{Strohmayer} \& {Watts}(2005)}]{Strohmayer2005}
{Strohmayer}, T.~E. \& {Watts}, A.~L. 2005, \apjl, 632, L111

\bibitem[{{Strohmayer} \& {Watts}(2006)}]{Strohmayer2006}
{Strohmayer}, T.~E. \& {Watts}, A.~L. 2006, \apj, 653, 593

\bibitem[{{Thompson} \& {Duncan}(1995)}]{Thompson1995}
{Thompson}, C. \& {Duncan}, R.~C. 1995, \mnras, 275, 255

\bibitem[{{van Hoven} \& {Levin}(2011)}]{vanHoven2011}
{van Hoven}, M. \& {Levin}, Y. 2011, \mnras, 410, 1036

\bibitem[{{van Hoven} \& {Levin}(2012)}]{vanHoven2012}
{van Hoven}, M. \& {Levin}, Y. 2012, \mnras, 420, 3035

\bibitem[{{Watts} \& {Strohmayer}(2006)}]{Watts2006}
{Watts}, A.~L. \& {Strohmayer}, T.~E. 2006, \apjl, 637, L117

\end{thebibliography}

\end{document}